\documentclass{emulateapj}
\usepackage{times}

\newcommand{\be}{\begin{equation}}
\newcommand{\ee}{\end{equation}}
\newcommand{\bd}{\begin{displaymath}}
\newcommand{\ed}{\end{displaymath}}
\newcommand{\bea}{\begin{eqnarray}}
\newcommand{\eea}{\end{eqnarray}}

\newcommand{\AAp}{\aap}
\newcommand{\ApJ}{\apj}

\newcommand{\km}{\,\hbox{km}}
\newcommand{\AU}{\,\hbox{AU}}

\newcommand{\sbs}[1]{\ensuremath{_\mathrm{#1}}}  

\renewcommand{\d}{\ensuremath{\mathrm{d}}}
\newcommand{\textmu}[1]{\micron}

\shortauthors{L{\"o}hne et al.}
\shorttitle{Decay of Debris Disks}
\slugcomment{Submitted to ApJ 1 Aug 2007, accepted 23 Oct 2007.}

\begin{document}


\title{Long-Term Collisional Evolution of Debris Disks}
\author{Torsten L{\"o}hne and Alexander V. Krivov}
\affil{Astrophysikalisches Institut und Universit{\"a}tssternwarte,
       Friedrich Schiller University Jena,
       Schillerg{\"a}{\ss}chen~ 2--3, 07745 Jena, Germany;
       tloehne@astro.uni-jena.de, krivov@astro.uni-jena.de}

\author{Jens Rodmann}
\affil{SCI-SA, Research and Scientific Support Department of ESA,
       ESTEC, 2200 AG Noordwijk, The Netherlands;
       jrodmann@rssd.esa.int
      }

\begin{abstract}
Infrared surveys indicate that the dust content in debris disks gradually
declines with stellar age. We simulated the long-term collisional depletion
of debris disks around solar-type (G2~V) stars with our collisional code.
The numerical results were supplemented by, and interpreted through, a new
analytic model. General scaling rules for the disk evolution are suggested.
The timescale of the collisional evolution is inversely proportional to the
initial disk mass and scales with radial distance as $r^{4.3}$ and with
eccentricities of planetesimals as $e^{-2.3}$. Further, we show that at actual
ages of debris disks between 10~Myr and 10~Gyr, the decay laws of the dust mass
and the total disk mass are different. The reason is that the
collisional lifetime of planetesimals is size-dependent. At any moment,
there exists a transitional size, which separates larger objects that still
retain the ``primordial'' size distribution set in the growth phase from smaller
objects whose size distribution is already set by disruptive collisions. The
dust mass and its decay rate evolve as that transition affects objects of
ever-larger sizes. Under standard assumptions, the dust mass, fractional
luminosity, and thermal fluxes all decrease as $t^\xi$ with $\xi = -0.3$...$-0.4$.
Specific decay laws of the total disk mass and the dust mass, including the
value of $\xi$, largely depend on a few model parameters, such as the critical
fragmentation energy as a function of size, the primordial size distribution
of largest planetesimals, as well as the characteristic eccentricity and
inclination of their orbits. With standard material prescriptions and a
distribution of disk masses and extents, a synthetic population of disks
generated with our analytic model agrees quite well with the observed
Spitzer/MIPS statistics of 24 and 70 \micron\ fluxes and colors versus age. 
\end{abstract}

\section{INTRODUCTION}

Since the IRAS discovery of the excess infrared emission around Vega by 
\citet{aumann-et-al-1984}, subsequent infrared surveys with ISO, Spitzer and other instruments 
have shown the Vega phenomenon to be common for main-sequence stars.
The observed excess is attributed to second-generation circumstellar dust,
produced in a collisional cascade from planetesimals and comets down to smallest
grains that are blown away by the stellar radiation. While the bulk of such a
debris disk's mass is hidden in invisible parent bodies,
the observed luminosity is dominated by small particles at dust sizes.
Hence the studies of dust emission offer a natural tool to gain insight into
the properties of planetesimal populations as well as planets that may shape them and,
ultimately, into the evolutionary history of circumstellar planetary systems.

In recent years, various photometric surveys of hundreds of nearby stars have
been conducted 
with the Spitzer Space Telescope.
These are the GTO survey of FGK stars
\citep{beichman-et-al-2005,bryden-et-al-2006,beichman-et-al-2006b},
the FEPS Legacy project  \citep{meyer-et-al-2004,kim-et-al-2005}, the A star GTO
programs
\citep{rieke-et-al-2005,su-et-al-2006}, the young cluster programs
\citep{gorlova-et-al-2006},
and others.
These observations were done mostly at 
24 and 70~\textmu{m} with the MIPS photometer, but also between 5 and
40~\textmu{m} with the IRS spectrometer 
\citep{jura-et-al-2004,chen-et-al-2006}.
Based on these studies,
about 15\% of mature solar-type (F0--K0) stars have been found to harbor cold
debris disks at 70~\textmu{m}. For cooler stars, the fraction drops to 0\%--4\%
\citep{beichman-et-al-2006b}.
For earlier spectral types, the proportion increases to about 33\%
\citep{su-et-al-2006}.
At 24~\textmu{m}, the fraction 
of systems with detected excess stays similar for A~stars, but appreciably decreases for FGK ones.
Similar results in the sub-millimeter range are expected to become available soon from a survey
with SCUBA and SCUBA2 on JCMT \citep{matthews-et-al-2007}.
Preliminary SCUBA results for M dwarfs suggest, in particular, that the
proportion of debris disks might actually be higher than suggested by Spitzer 
\citep{lestrade-et-al-2006}.

All authors point out a decay of the observed infrared excesses with systems'
age.
However, the values reported for the slope of the decay, assuming a
power-law dependence $t^{-\alpha}$, span a wide range. \citet{greaves-wyatt-2003}
suggest $\alpha \la 0.5$, \citet{liu-et-al-2004} give $0.5 < \alpha < 1.0$,
\citet{spangler-et-al-2001} report $\alpha \approx 1.8$, and
\citet{greaves-2005} and \citet{moor-et-al-2006} derive $\alpha \approx 1.0$.
Fits of the upper envelope of the distribution of luminosities over the age yield
$\alpha \approx 1.0$ as well \citep{rieke-et-al-2005}. Besides, the dust fractional
luminosity exhibits a large dispersion at any given age.

In an attempt to gain theoretical understanding of the observed evolution,
\citet{dominik-decin-2003} assumed that equally-sized ``comets''
produce dust through a cascade of
subsequent collisions among ever-smaller objects. If this dust is removed by
the same mechanism, the steady-state amount of dust in such a
system is proportional to the number of comets. This results in an
$M/M_0 \approx \tau/t$
dependence for the amount of dust and for the number of
comets or the total mass of the disk.
Under the assumption of a steady state, this result is valid even for more complex
systems with continuous size distributions from planetesimals to dust.
Tenuous disks, where the lifetime of dust grains is not
limited by collisions but by transport processes like the Poynting-Robertson
drag \citep{artymowicz-1997,krivov-et-al-2000,wyatt-2005}, follow $M \propto t^{-2}$
rather than $M \propto t^{-1}$.

More recently, \citet{wyatt-et-al-2007a} lifted the most severe simplifying
assumption
of the Dominik-Decin model, that of equal-sized parent bodies, and included them
into the collisional cascade. A debris disk they consider is no longer a two-component
system ``comets + dust''. Instead, it is a population of solids with a continuous
size distribution, from planetesimals down to dust.
A key parameter of the description by \citet{dominik-decin-2003} is the
collisional lifetime of comets, $\tau$.
\citet{wyatt-et-al-2007a} replaced it with the lifetime of the largest
planetesimals
and worked out the dependencies on this parameter in great detail.
Since the collisional timescale is inversely proportional to the amount 
of material, $\tau \propto 1/M_0$, the asymptotic disk mass becomes independent of its initial mass.
Only dynamical quantities, i.e. the disk's radial position and extent, the orbiting objects' eccentricities 
and inclinations, and material properties, i.e. the critical specific energy and the 
disruption threshold, as well as the type of the central star determine the very-long-term 
evolution.

Still, there are two important simplifications made in the model by
\citet{wyatt-et-al-2007a}:
(i) the disk is assumed to be in collisional equilibrium at all sizes, from dust up to the 
largest planetesimals 
and (ii)
the minimum specific energy needed to disrupt colliding objects
is independent of their size. As a consequence of (i) and (ii),
the size distribution of solids is a single power-law.
To check how reasonable these assumptions are, realistic simulations of the disks with 
collisional codes are necessary
\citep[e.g.,][]{thebault-et-al-2003,krivov-et-al-2005,krivov-et-al-2006,thebault-augereau-2007}.

The aim of this paper is two-fold. First, we follow the evolution of debris disks
with our elaborate numerical code  \citep{krivov-et-al-2005,krivov-et-al-2006}
to check the existing analytic models and the assumptions (i) and (ii) they are based upon.
Second, in order to make these numerical results easier to use,
we develop a new analytic model for the evolution of disk mass and dust mass
that relaxes both assumptions (i) and (ii) above.

Section~\ref{secNumerics}
summarizes the basic ideas and assumptions and describes our numerical model
and the runs of the collisional code. In Section~\ref{secScalings} the numerical results
are presented and dependences of the collisional timescale on the disk mass, distance to the 
star, and mean eccentricity of parent bodies are derived.
In section~\ref{secAnalytics}, the analytic model
for the evolution of disk mass and dust mass is developed.
Section~\ref{secEvolLuminosity} analyzes the evolution of dust luminosities.
In Section~\ref{secObservations}, we use the analytic model to synthesize representative
populations of debris disks and compare them with statistics of debris disks
derived from the Spitzer surveys.
A summary is given and conclusions are drawn in Section~\ref{secConclusions}.
\pagebreak

\section{NUMERICAL MODEL AND DESCRIPTION OF RUNS}
\label{secNumerics}
\subsection{Basic Approach}

For all numerical runs in this paper, we use a C++-based collisional
code (ACE, Analysis of Collisional Evolution).
The code numerically solves the Boltzmann-Smoluchowski kinetic equation
to evolve a disk of solids in a broad range of sizes (from 
sub-micrometers to about a hundred of kilometers),
orbiting a primary in nearly-Keplerian orbits
(gravity + direct radiation pressure + drag forces) and
experiencing disruptive collisions.
Collisions are simulated with available material- and size-dependent scaling
laws for fragmentation and dispersal in both strength and gravity regime.
The current version implements a 3-dimensional kinetic model, with masses,
semi-major axes, and eccentricities as phase space variables.
This approach automatically enables a study of the simultaneous evolution of
mass, spatial, and velocity distribution of particles.
The code is fast enough to easily follow the evolution of a debris disk over
Gyr timescales.
A detailed description of our approach, its numerical implementation, and astrophysical
applications can be found in our previous papers
\citep{krivov-et-al-2000,krivov-et-al-2005,krivov-et-al-2006}.

\subsection{Disruption Threshold and Critical Specific Energy}
\label{secQD}

An object is said to be disrupted in a collision, if the largest fragment is at
most
half as massive as the original object. If the impactor's relative velocity is so
high that the ratio of impact energy and target mass exceeds the
target's critical specific energy, $Q\sbs{D}^*$, the target (and the impactor)
are disrupted.
For small objects, this binding energy is dominated by material
strength,
and for larger objects, self-gravity takes
over.
Both regimes are usually described by a sum of two power laws
\citep[][Sect.~5.1, and references therein]{krivov-et-al-2005}
\begin{equation}
  Q\sbs{D}^* = A\sbs{s} \left(\frac{s}{1\:\mathrm{m}}\right)^{3b\sbs{s}} +
                A\sbs{g} \left(\frac{s}{1\:\mathrm{km}}\right)^{3b\sbs{g}},
           \label{eqQD}
\end{equation}
where ``s'' and ``g'' stand for strength and gravity, respectively.
The reported values of the coefficients $A\sbs{s}$ and $A\sbs{g}$ vary by more
than one order of magnitude, and we took $A\sbs{s} = A\sbs{g} = 5 \times 10^6$~erg/g,
in agreement with the reference case for
basalt given by \citet{benz-asphaug-1999}. The
exponents are $3b\sbs{s} = -0.3$ and $3b\sbs{g} = 1.5$ (corresponding to $-0.1$ and $0.5$
in the mass scaling). With these
parameters, the two power-law components contribute equally at $s \approx
316$~m, and the lowest binding energy, the minimum $Q\sbs{D}^*$, is reached
at $s \approx 129$~m. The influence of the choice of parameters on the
resulting evolution is discussed in Sect.~\ref{secAnalytics}.

For computational reasons, we refrained from including a treatment of cratering
collisions in the runs.
Note that these were not taken into account in previous studies of the long-term
evolution of debris disks \citep[e.g.][]{dominik-decin-2003,wyatt-et-al-2007a}
either.
\citet{thebault-et-al-2003} and
\citet{thebault-augereau-2007}, who focused on shorter time spans, did include this
non-disruptive type of collisions that lead to the continuous
erosion of a target by small impacting projectiles. They found the effect to
be dominant for particles in between 100~\textmu{m} and 1~cm for the case
of the inner $\beta$ Pictoris disk, while big, kilometer-sized objects in the
gravity regime are mainly lost to disruptive collisions \citep[see Table 4 in][]{thebault-et-al-2003}.
However, 
including cratering
can lower the lifetime of large objects, especially when relative
velocities are low and disruptive collisions are rare.
Another caveat is that cratering collisions alter the shape of the wavy size distribution 
at the lower end \citep{thebault-augereau-2007}, which
affects the observable thermal fluxes.

\subsection{Collisional Outcomes}

The distribution of sizes and the velocities of fragments in an
individual (catastrophic) collision has been subject to studies for decades.
Laboratory work was done on high-velocity impacts on scales of millimeters and
centimeters
\citep[e.g.,][]{fujiwara-et-al-1977,fujiwara-1986,davis-ryan-1990}.
Statistics on the mass distributions of observed asteroidal families
and geometrical considerations
\citep{paolicchi-et-al-1996,tanga-et-al-1999,tedesco-et-al-2005} as well as
gravito-hydrodynamic simulations of fragmentation and reaccumulation
\citep{michel-et-al-2002} cover the range of larger, kilometer-sized
bodies.
On small scales, the resulting size distributions show a strong dependence
on impact velocity and seem to indicate a turn in the power law at
fragment sizes around $\approx 1$~mm (or $\approx 1$\%\ of the size of the
used targets). The slope for objects above that size is steeper than the one
for smaller objects \citep{davis-ryan-1990}. However, \citet{thebault-et-al-2003}
found that the ratio of these two slopes and the size at which
the slope changes influence simulation results only slightly.
On kilometer and larger scales, the fragmentation is influenced by gravitational
reaccumulation of relatively small fragments onto bigger ones. Hence, bigger
fragments ($\sim 100$km) will be overabundant, and conversely, smaller fragments
($\sim 1$km) underabundant, compared to the underlying distribution without gravity.
The slopes of the size distribution $n(s)\propto s^{-p}$ of kilometer-sized objects
are poorly known. A wide range from $p=3.5$ up to $p=9.0$ has been reported.
These deviations in the kilometer regime are most probably the severest caveat
of the power-law approximation, because they are independent of
the actual material and caused only by gravity.
Nevertheless, we assume that fragments follow a single
power-law distribution $n\sbs{frag} (s) \propto s^{-3.5}$, expecting the
influence on the final collisional steady state to be only moderate.

\subsection{Commons for All Runs}
\label{secCommons}

All disk models presented here are set up around a star of solar mass and
luminosity. Parameters of the central star affect the disk evolution in
various ways. They determine the size limit for grain's blowout by
radiation pressure and orbital velocities at a given
distance, thereby altering impact velocities and rates.
For late-type stars, strong stellar winds may affect the dust dynamics
\citep{augereau-beust-2006,strubbe-chiang-2006}.
On the observational
side, dust temperatures and brightnesses are influenced.
Here, we focus on the scalings for a fixed spectral type (G2V),
and not on scalings between different types.

The disks themselves all share the same material properties and shapes.
We adopt the material, described by a bulk density $\rho=2.5$~g/cm$^3$, the
radiation pressure efficiency of astronomical silicate
\citep{laor-draine-1993}, and a
critical fragmentation energy as specified in Sect.~\ref{secQD}.
We switched off the Poynting-Robertson effect, which is unimportant for debris disks
under study, as well as stellar wind drag, which plays only a minor role
around G-type stars.
The fragments produced in an individual collision are distributed according
to a single power law,
$\d N\propto s^{-3.5} \d s \propto m^{-11/6} \d m$.
A biggest fragment size is assumed to scale with
specific impact energy to the power of $1.24$ \citep[for details,
see][]{krivov-et-al-2006}.
The initial mass distribution is given
by $\d N \propto m^{-q}$, with $q = 1.87$, a value that accounts for the
modification of the classical Dohnanyi's (\citeyear{dohnanyi-1969})
$q=1.833$ through the size dependence of material
strength \citep[see, e.g.,][]{durda-dermott-1997}. The particle masses range
from $4.2\times 10^{-15}$~g, corresponding to a radius of 74~nm, to
$4.2\times 10^{21}$~g, corresponding to 74~km. The stepping between the 60 mass
bins is logarithmic with a factor of $\approx 4$ between neighboring bins.
The initial radial profile of the particle density was given by a
slope of the normal optical depth of $-1.0$. 
The initial total mass of each disk was set to 
1~$M_\oplus$ (earth mass).

\subsection{Specifics of Individual Runs}

\notetoeditor{The table should use only one column in the final version.}
\begin{deluxetable}{rcc}
\tablecaption{Description of numerical runs.\label{tab_runs}}
\tablewidth{0pt}
\tablehead{\colhead{Run} & \colhead{Distance [AU]} & e\sbs{max}}
\startdata
\multicolumn{3}{c}{\em Nominal runs}\\
ii-0.3 & 7.5--15 & 0.3 \\
 i-0.3 & 15--30  & 0.3 \\
 o-0.3 & 30--60  & 0.3 \\
oo-0.3 & 60--120 & 0.3 \\
\multicolumn{3}{c}{\em Additional runs}\\
 i-0.1 & 15--30  & 0.1 \\
 i-0.2 & 15--30  & 0.2 \\
 i-0.4 & 15--30  & 0.4 \\
\enddata
\end{deluxetable}

We have made four ``nominal'' runs, each of which corresponds to a certain radial
part of the disk between 7.5 and 120 AU from the star (Table~\ref{tab_runs}).
In these runs we assumed initial eccentricities of planetesimals
to be uniformly distributed between $e\sbs{min} = 0.0$
and $e\sbs{max} = 0.3$, spanning three bins centered at 0.05, 0.15 and 0.25.
In addition, three runs with altered maximum eccentricity of 0.1,
0.2, and 0.4 were made for the $15$--$30\AU$ ring.
In all the runs, we assumed that orbital inclinations are distributed between
$I\sbs{min} = e\sbs{min}/2$ and $I\sbs{max} = e\sbs{max}/2$
in accordance with the energy equipartition relation $I = e/2$.

\section{NUMERICAL RESULTS AND SCALING LAWS}
\label{secScalings}

\subsection{Evolution of Disks of Different Masses}

A debris disk is said to be in a quasi-steady state or quasi-equilibrium
if the amounts of particles with different sizes on different orbits,
while changing with time (therefore ``quasi''), stay constant relative to each other.
For brevity, we will often omit ``quasi'' and use simply
``steady state'' or ``equilibrium''.
To express the condition of a quasi-steady state
formally, we can introduce a phase space, in which
a dynamical state of each particle is characterized by a vector {\bf p}.
That vector may be composed, for instance, of coordinates
and velocity components. Alternatively, {\bf p} may represent the
set of orbital elements of the object.
Let $n({\bf p}, s, t)$ be the number of objects with radii in $[s, s + \d s]$
at phase space ``positions'' $[{\bf p}, {\bf p} + \d {\bf p}]$ that the disk contains
at the time instant $t$.
The assumption of a quasi-steady state can now be expressed as
\begin{equation}
  n({\bf p}, s, t) = \tilde{n}({\bf p}, s) \: f(t) .
\label{ntilde}
\end{equation}
The total disk mass,
\begin{equation}
  M\sbs{disk} (t) = \int\int\nolimits n({\bf p}, s, t) \d {\bf p} \d s,
 \label{eqMDef}
\end{equation}
can be rewritten as
\begin{equation}
  M\sbs{disk} (t) = f(t) \int\int\nolimits \tilde{n}({\bf p}, s) \d {\bf p} \d s
\end{equation}
or, setting $f(0) = 1$,
\begin{equation}
  M\sbs{disk} (t) = f(t) M_0 \label{eqMdisk},
\end{equation}
where $M_0$ is the initial disk mass.
As long as objects are both created and lost in two-particle collisions,
their gain and loss rates are given by
\begin{eqnarray}
  \dot{n} ({\bf p}, s, t) &=&
  \int\int\int\int\nolimits
\left[G({\bf p}, s, {\bf p}_1, s_1, {\bf p}_2, s_2)
\right.
\nonumber\\
&&
\left.
      -
      L({\bf p}_1, s_1, {\bf p}_2, s_2)
      \delta({\bf p} - {\bf p}_1)
      \delta(s - s_1)
\right]
\nonumber\\&\times&
\tilde{n}({\bf p}_1, s_1) \: f(t) \:
\tilde{n}({\bf p}_2, s_2) \: f(t)
\nonumber\\&\times&
\d {\bf p}_1 \d s_1 \d {\bf p}_2 \d s_2,
\label{eqndot}
\end{eqnarray}
where
the function $G({\bf p}, s, {\bf p}_1, s_1, {\bf p}_2, s_2)$
describes the gain in population $\bf p$, $s$ due to collisions
between ${\bf p}_1$, $s_1$ and ${\bf p}_2$, $s_2$
and the function $L({\bf p}_1, s_1, {\bf p}_2, s_2)$
accounts for the loss in population ${\bf p}_1$, $s_1$ 
in collisions with ${\bf p}_2$, $s_2$.
The disk mass changes at a rate
\begin{equation}
  \dot{M}\sbs{disk} (t) = \int\int\nolimits \dot{n}({\bf p}, s, t)
                          \d {\bf p} \d s \label{eqMdot1}
\end{equation}
or
\begin{equation}
  \dot{M}\sbs{disk} (t) = \dot{f}(t) \int\int\nolimits \tilde{n}({\bf p}, s)
                          \d {\bf p} \d s \label{eqMdot2}.
\end{equation}
From Eqs.~(\ref{eqndot}) and (\ref{eqMdot1}), we find that
$\dot{M}\sbs{disk}(t) \propto f^2(t)$, while Eq.~(\ref{eqMdot2}) suggests
$\dot{M}\sbs{disk}(t) \propto \dot{f}(t)$. Hence, $\dot{f}(t)\propto f^2(t)$.
Integration yields
\begin{equation}
  f = \frac{1}{1+t/\tau} .
\label{dd}
\end{equation}
Using Eq.~(\ref{eqMdisk}) we obtain
\begin{equation}
  M\sbs{disk} (t) = \frac{M_0}{1+t/\tau} \label{eqMOverTime}
\end{equation}
and
\begin{equation}
  \dot{M} \sbs{disk} (t) = - C M\sbs{disk}^2, \label{eqCdef}
\end{equation}
where
$1/C = M_0 \cdot \tau$, i.e. the product of the initial mass and a
characteristic time.
This relation is invariant under the transformation
$(t,M\sbs{disk})\rightarrow(t\cdot x, M\sbs{disk}/x)$, even if $C$ is not constant.
Therefore, the mass scale of a system under collisional evolution is
inversely proportional to its timescale. For example, doubling the initial total mass
halves the collisional lifetime of the system.
All curves in the $M\sbs{disk}(t)$ plots can be shifted along lines of equal $t \cdot 
M\sbs{disk}$.

\citet{dominik-decin-2003} used this approach and equated the characteristic
time $\tau$ with the collisional lifetime of their ``comets''.
At the initial phase $t\ll\tau$, Eq. (\ref{eqMOverTime}) gives
\begin{equation}
  M\sbs{disk}(t) \approx M_0 \left( 1 - t/\tau \right).
  \label{eqMOverTimeApprox}
\end{equation}
If the system is old enough so that $t\gg\tau$, the total mass will be
just proportional to $t^{-1}$.
Particles whose lifetimes are independent of the total mass are
exempt from the asymptotic one-over-$t$ behavior. Examples would be the
$\beta$-meteoroids that are blown out and small particles in disks tenuous
enough for the Poynting-Robertson effect to be more
efficient than collisions. The total mass of such particles is
$\propto t^{-2}$ \citep{dominik-decin-2003}.

\begin{figure}[t!]
  \includegraphics{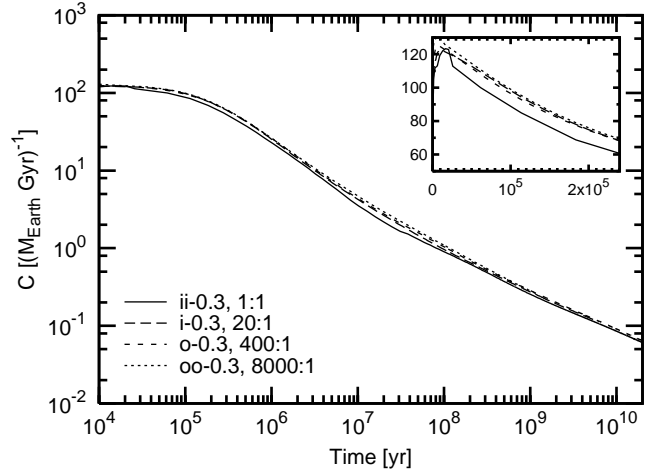}\\
  \caption{The coefficient $C$ from Eq. (\ref{eqCdef}) as a function of time for four
  nominal runs. The total disk mass and
  time in the runs are scaled according to $M\sbs{disk}\propto t^{-1}$
  to compensate for the difference in dynamical timescale.
  Note that the near-constancy of $C$ at the beginning of the evolution is an
  artefact of the double-logarithmic plotting. The double-linear inset shows
  that the decrease of $C$ is {\em fastest} at earlier times.
\label{figCOverTime}}
\end{figure}

As we have shown, for the systems that undergo a steady-state collisional evolution,
the factor $C$ in Eq.~(\ref{eqCdef}) (or $\tau$) should be constant.
To check this, we evaluated $C = -\dot M\sbs{disk}/M\sbs{disk}^2$ for every
two subsequent time steps of the numerical runs.
The results are given in Fig. \ref{figCOverTime}.

\begin{figure}[t!]
  \includegraphics{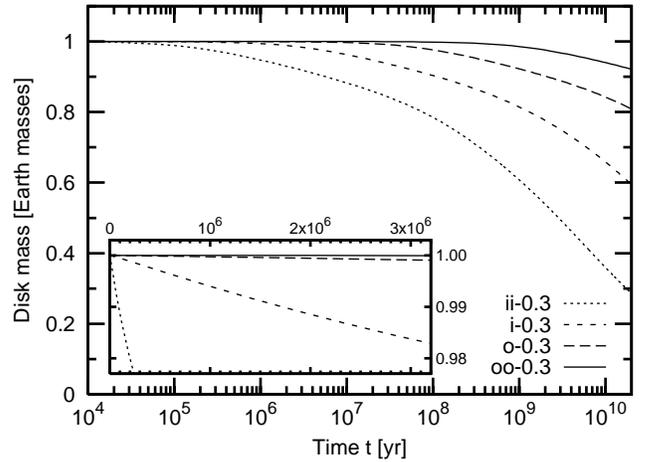}\\
  \caption{The evolution of the total mass in the four nominal runs. 
     Again, the plateau at the beginning of the evolution is an artefact of the
     logarithmic plotting of time. In fact, the mass decay is strongest at the
     very beginning (see inset and Eqs.~(\ref{eqMOverTime}), (\ref{eqMOverTimeApprox})).
  \label{figMOverTime}}
\end{figure}

Instead of being constant at later times, $C$ decreases, roughly following a
power law $C\propto t^{-2/3\ldots-4/5}$.
The explanation is simple: the systems did not reach an equilibrium
where $t\gg\tau$ or at least $t\approx\tau$ during their lifetime.
The evolution of the total mass in Fig.~\ref{figMOverTime} demonstrates that as
well.

\subsection{Dependence on Distance from the Star}

Rings of identical mass but at different distances have different
collisional timescales. The comparison in Fig. \ref{figCOverTime} shows that
doubling the distance requires a 20-fold increase in disk mass to have the
same timescale. This corresponds to a power-law dependence
\be
  C \propto r^{-4.3}.\label{eqScalingR}
\ee
In a thorough analytic approach based on a Dohnanyi-type collisional cascade,
\citet{wyatt-et-al-2007a} came up with
$C \propto r^{-13/3}$, which is in good agreement with our numerical result.
This index is made up of three contributions. First, the density in the rings
drops with $r^{-3}$ as their circumference, height, and width increase
linearly. Second, the relative velocities have an $r^{-1/2}$ dependence.
Third, these impact velocities affect the minimum required mass for a projectile to be
disruptive and thereby the total number of such projectiles. That gives
another $r^{1-q}$, where $q$ is the slope in the
appropriate mass distribution, e.g. $q = 11/6$ for the classical Dohnanyi case.
See Sect.~\ref{secCollTime} for details.

\subsection{Dependence on Eccentricities of Parent Bodies}
\label{secEcc}

\begin{figure}[t!]
  \includegraphics{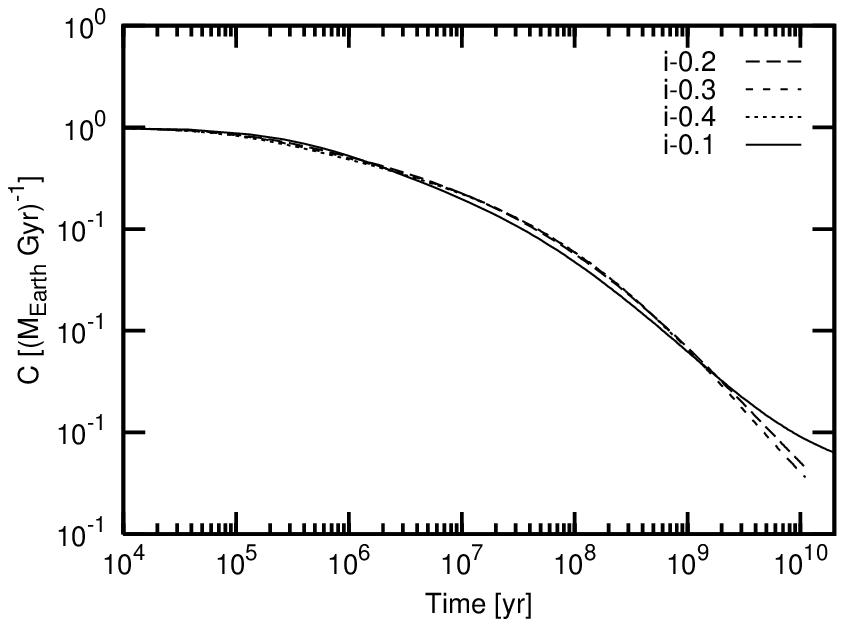}\\
  \includegraphics{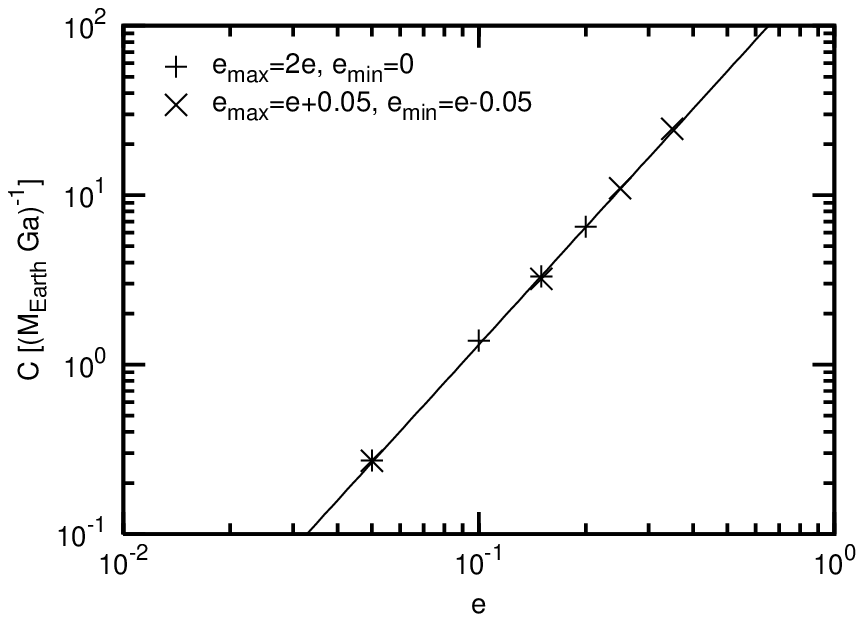}\\
  \caption{The influence of the average eccentricity of planetesimals 
  on the timescale of disk's collisional evolution.
  \textsl{Top:} the evolution of the parameter $C$ from
  Eq.~(\ref{eqCdef}) for four different runs (i-0.1, \ldots, i-0.4).
  \textsl{Bottom:}
  four initial $C$ values versus average eccentricity $e = (e\sbs{max} +
  e\sbs{min})/2$ (pluses)  together with the $C \propto e^{9/4}$ fit for those runs
  (line) and the same for runs with a narrower range of eccentricities,
  as described in Sect.~\ref{secEcc} (crosses).
  \label{figCOverEMax}}
\end{figure}

The intrinsic collisional probability of planetesimals is nearly independent
of their eccentricities, as long as they are not too high \citep[see, e.g.][]{krivov-et-al-2006}.
Nevertheless, eccentricities determine impact velocities and, through that, the
minimum size of a disruptive projectile. Therefore, higher planetesimal eccentricities
imply a larger rate of catastrophic collisions and thus a faster collisional
evolution.
To quantify the dependence, we have made runs with maximum eccentricities
of 0.1, 0.2, 0.3, and 0.4 (Table~\ref{tab_runs}) and determined the
values of $C$.
The results suggest a power law $C \propto e\sbs{max}^{9/4}$ as shown in
Fig.~\ref{figCOverEMax}.

This result comes as a surprise. \citet{wyatt-et-al-2007a} derive $C \propto
e^{5/3}$.
The same scaling is inherited by our analytic model, see
Eq.~(\ref{eqDependencies}) below.
Since this discrepancy can be either due to an incompleteness of the analytic
approach or to a non-linear relation between the maximum and the effective
eccentricity, we tried to rule out the latter case by
performing additional runs with $e$ confined to narrow bins of
width $0.1$, centered at $0.05$, $0.15$, $0.25$, and $0.35$.
These runs can be well described by the same power law,
$C \propto e^{9/4}$ (Fig.~\ref{figCOverEMax}).
Therefore, the analytic model fails to reproduce this particular dependence.
Nevertheless, it correctly describes many others, as the next sections will 
show.

\section{ANALYTIC MODEL FOR EVOLUTION OF DISK MASS AND DUST MASS}
\label{secAnalytics}

\subsection{Size and Mass Distributions}

In what follows, we will analyze size or mass distributions of objects.
Different authors use distributions of different physical quantities
(number, cross section, mass) with different arguments (particles size or mass)
and of different type (differential, cumulative, per size decade, etc.).
A standard choice is to use a differential size distribution,
$n(s)$, that gives the number of particles per unit size interval:
\begin{equation}
  n(s) \equiv \int\nolimits n({\bf p}, s) \d {\bf p} ,
\end{equation}
or a differential mass distribution, $n(m)$, that gives the number of particles
per unit mass interval.
Instead of $n$, it is often convenient to use the mass-per-size-decade
distribution,
\begin{equation}
  \frac{\d M\sbs{disk}}{\d \log_{10} s} = \ln (10) \: s\:m(s)\:n(s) .
\end{equation}
In contrast to $n(s)$, this quantity tells us directly, objects in which size range
contribute the most to the mass of the system.
Therefore, we will use it when plotting size or mass distributions.

In the case of a power-law size distribution,
$n(s) \d s \propto s^{2-3q} \d s$
is the number of objects with sizes $[s, s+\d s]$
and $n(m) \d m \propto m^{-q} \d m$
is the number of objects with masses $[m, m+\d m]$.
The mass per size decade is $\propto s^{6-3q} \propto m^{2-q}$.
When $q < 2$, the total mass is determined by large bodies,
whereas the cross section is dominated by small particles as long as $q > 5/3$.

\subsection{Three-Slope Distribution}

\label{sec3slope}

\begin{figure}[t!]
\includegraphics{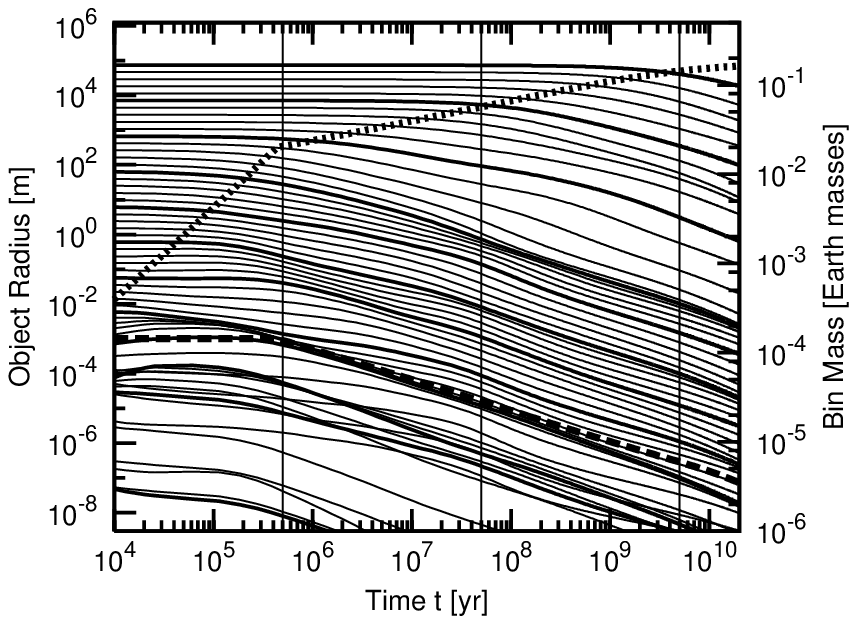}\\
\includegraphics{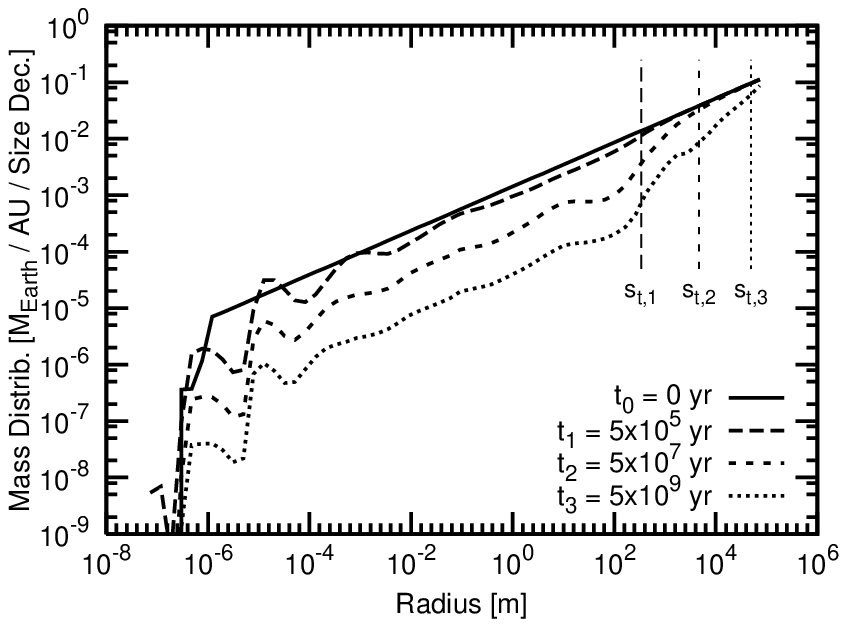}\\
 \caption{
           Results of the ii-0.3 run.
           {\em Top:}
           Time evolution of mass in individual mass bins, from the 
           largest bodies of $74$~km in radius to the smallest, $74$~nm in
           radius. The mass ratio between adjacent bins is 4.
           Each solid line corresponds to one individual bin and gives
           the mass contained in that bin (see the right axis) as a function of
           time.
           The left axis can be used to find the line that corresponds to a
           given object size. 
           The thick dashed curve corresponds to $\approx 1$~mm radius, i.e. to
           the largest solids still treated as dust. 
           The thick dotted curve, which goes roughly through the turning points
           of the curves, is the transition size $s\sbs{t} (t)$; see
           Eq.~(\ref{eqEquiMass}).
           {\em Bottom:}
           Size/mass distribution at four specific instants of time
           shown in the top panel with vertical lines:
           initially,
           after $5\times 10^5$ years when $s\sbs{t}$ has reached $s\sbs{b}$,
           and after $5\times 10^7$ and $5\times 10^9$~yr when significant
           dust depletion has already occurred.
  \label{figMassDistribution}}
\end{figure}

The combination of material strength at smaller
sizes and self-gravity at larger ones, with a turnover at around 100~m,
causes the size distribution in a collisionally evolving system to strongly deviate
from a single-slope power law,
especially for object sizes of around 1~km.
This is illustrated by Fig.~\ref{figMassDistribution}
that shows how  a disk evolves from the first-guess power
law to a more realistic size distribution.
The speed of this evolution is determined by the collisional timescales of
populations of different-sized particles
in the disk. Populations of smaller particles with sufficiently short lifetimes
consist mostly of fragments of disruption of larger bodies. They will have 
reached collisional equilibrium with each other soon, according to their production rate by
populations with longer lifetimes. Those latter populations of bigger
particles will still be on their way to a steady state. As time goes by,
more and more long-lived populations will undergo the transition from
primordial to reprocessed material.

As this transitional mass moves towards larger objects with time,
the smaller particles follow to a new ``intermediate steady state''.
The lower panel of Fig.~\ref{figMassDistribution} shows the development of the
characteristic wavy shape in the size distribution
\citep[e.g.,][]{campo-bagatin-et-al-1994,thebault-et-al-2003,krivov-et-al-2006}
at the small-size end near the blowout limit due to radiation pressure.
Once established, this shape remains constant. Only the absolute level
changes because this distribution at smaller sizes acts as the trail of the
distribution at larger sizes.
In the upper panel of Fig.~\ref{figMassDistribution}, the number
of smaller particles is constant for some time and then
goes down,
as soon as the distribution in the gravity regime starts to deviate
from its primordial one.

\begin{figure}[t]
  \includegraphics{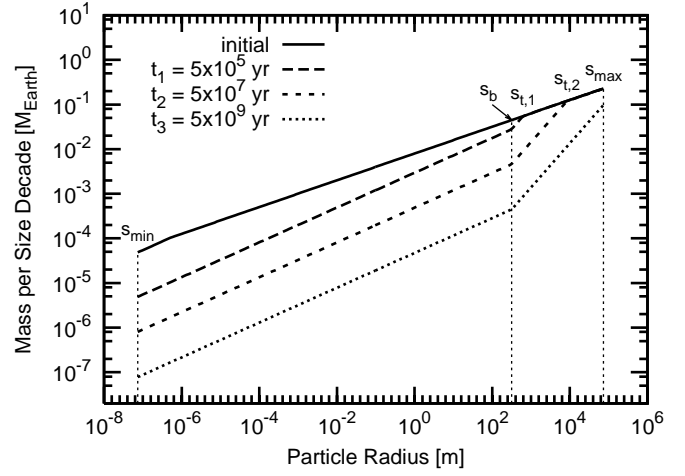}\\
  \caption{Schematic plot of the three regimes in the mass distribution
    and its time evolution.
    The mass $s\sbs{t}$ divides second generation material in collisional
    equilibrium ($s < s\sbs{t}$) from primordial material ($s > s\sbs{t}$),
    while $s\sbs{b}$ divides the material strength regime ($s < s\sbs{b}$) from
    the gravity regime ($s > s\sbs{b}$).
  \label{figMassDistScheme}}
\end{figure}
These arguments suggest that an overall size distribution $n(s)$ can be
approximated by a combination of
three power laws (Fig. \ref{figMassDistScheme}).
For particles large enough to be only barely affected by collisions at time $t$,
we assume $n$ to follow $s^{2-3q\sbs{p}}$.
Here, $q\sbs{p}$ is the ``primordial'' slope determined by the processes in which these
planetesimals have formed.
Small particles that are in quasi-steady state are separated from bigger
primordial objects by a transition zone which we characterize by a
time-dependent size $s\sbs{t}(t)$.
To distinguish between the strength and gravity regimes,
we introduce two more power laws and assume the mass distribution
to follow $n\propto s^{2-3q\sbs{g}}$ for gravity-dominated quasi-steady state
and $n\propto s^{2-3q\sbs{s}}$ for strength-dominated quasi-steady state.
The two regimes are separated by an object size $s\sbs{b}$, which we
will call breaking radius. Thus, the waviness is neglected, but 
the effect of a size-dependent $Q\sbs{D}^*$ is kept.

The resulting size distribution is given by
\begin{equation}
  n(s) = n\sbs{max}\left(\frac{s\sbs{max}}{s}\right)^{3q\sbs{p}-2} \label{eqDist0}
\end{equation}
for $s\sbs{t} \leq s < s\sbs{max}$,
\begin{equation}
  n(s) = n\sbs{max}\left(\frac{s\sbs{max}}{s\sbs{t}}\right)^{3q\sbs{p}-2}
                  \left(\frac{s\sbs{t}}{s}\right)^{3q\sbs{g}-2}
\end{equation}
for $s\sbs{b} \leq s < s\sbs{t}$, and
\begin{equation}
  n(s) = n\sbs{max}\left(\frac{s\sbs{max}}{s\sbs{t}}\right)^{3q\sbs{p} - 2}
                  \left(\frac{s\sbs{t}}{s\sbs{b}}\right)^{3q\sbs{g} - 2}
                  \left(\frac{s\sbs{b}}{s}\right)^{3q\sbs{s} - 2}\label{eqnm}
\end{equation}
for $s\sbs{min} < s < s\sbs{b}$,
where $n\sbs{max} \equiv n(s\sbs{max})$,
with $s\sbs{max}$ being the size of the largest planetesimals.
From this distribution, two important quantities can be derived. One is the
total {\em disk} mass,
\begin{equation}
  M\sbs{disk} = \int_{s\sbs{min}}^{s\sbs{max}}\limits n(s)
   \frac{4}{3}\pi\rho s^3 \d s,
  \label{eqTotalMass}
\end{equation}
and the other is {\em dust} mass (that determines the infrared luminosity
and therefore provides a link to observations),
\begin{equation}
  M\sbs{dust} = \int_{s\sbs{min}}^{s\sbs{d}}\limits n(s)
   \frac{4}{3}\pi\rho s^3 \d s,
  \label{eqDustMass}
\end{equation}
where $s\sbs{min} \leq s\sbs{d} < s\sbs{b}$.

\subsection{Collisional Lifetimes of Planetesimals}
\label{secCollTime}

As seen from Eqs.~(\ref{eqDist0})--(\ref{eqDustMass}),
the evolution of $M\sbs{disk}$ and $M\sbs{dust}$ is controlled by
$n\sbs{max} (t)$ and $s\sbs{t} (t)$.

We start with $n\sbs{max}$
and assume, according to Eqs.~(\ref{ntilde}) and~(\ref{dd}):
\begin{equation}
  n\sbs{max}(t) = \frac{n\sbs{max}(0)}{1+t/\tau\sbs{max}},
\label{DD}
\end{equation}
where $\tau\sbs{max}$ is the  collisional lifetime of these largest bodies.
Equation~(\ref{DD}) closely reproduces the disk evolution as soon as the whole system has
reached the quasi-steady state
at all sizes or, in other words, as soon as
$s\sbs{t}(t)$ has reached $s\sbs{max}$.

The second quantity that we need, $s\sbs{t} (t)$, could easily be obtained
by inverting the function $\tau(s)$, the collisional lifetime of planetesimals
of a given size $s$. To obtain $\tau(s)$, we begin with the lifetime
of the largest objects in a disk.
Assuming that $q > 5/3$,
\citet[][their Eq.~12]{wyatt-et-al-2007a} approximated it as
\begin{eqnarray}
  \tau\sbs{max} &=& \frac{4\pi}{\sigma\sbs{tot}}\cdot
    \left(\frac{s\sbs{max}}{s\sbs{min}}\right)^{3q\sbs{p}-5}
    \!\!\!\!\!\!
    \cdot \frac{r^{5/2}\d r}{\left(\mathsf{G} M_*\right)^{1/2}}
    \nonumber\\
    &\times& \frac{I}{f(e,I) G(q, s)}.
    \label{eqtc}
\end{eqnarray}
where $e$ and $I$ are the effective orbital eccentricities and
inclinations, $\sigma\sbs{tot}$ is the initial cross sectional area of the disk
material, $\mathsf{G}$ the gravitational constant, $r$ the radial distance of
the ring of parent bodies, and $\d r$ its width. The slope $q$ in their
single-power-law approach corresponds to the primordial slope $q\sbs{p}$ in our
nomenclature. The functions $f$ and $G$ are given by
\begin{eqnarray}
  f(e,I) &=& \sqrt{\frac{5}{4}e^2 + I^2},\\
  G(q,s) &=& \left[ X\sbs{c}(s)^{5-3q}
                    - \left(\frac{s\sbs{max}}{s}\right)^{5-3q} \right]\nonumber\\
                &+& 2\frac{q-5/3}{q-4/3} \left[ X\sbs{c}(s)^{4-3q}
                    - \left(\frac{s\sbs{max}}{s}\right)^{4-3q} \right]
                \nonumber\\
                &+& \frac{q-5/3}{q-1}
                       \left[ X\sbs{c}(s)^{3-3q}
                    - \left(\frac{s\sbs{max}}{s}\right)^{3-3q} \right],\label{eqGFull}
\end{eqnarray}
with
\begin{eqnarray}
  X\sbs{c}(s) &=& \left(
                \frac{2Q\sbs{D}^*(s) \; rf(e,I)^{-2}}{\mathsf{G}M_*}
               \right)^{1/3}.\label{eqXcFull}
\end{eqnarray}
While $f(e,I)$ describes the dependence of the impact velocities on
eccentricities and inclinations, the functions $G$ and $X\sbs{c}$ characterize
the disruption of planetesimals by smaller projectiles.
Namely, $X\sbs{c}(s)$ is the minimum size ratio between the smallest disruptive projectile
and the target, and $G(q,s)$ is the number of disruptive projectiles.

We need the lifetime of objects of an arbitrary size, $\tau(s < s\sbs{max})$.
To derive it, we can simply substitute $s\sbs{max}$ by $s$ in
Eq.~(\ref{eqtc}), obtaining
\begin{eqnarray}
  \tau(s) &=& \frac{4\pi}{\sigma\sbs{tot}}\cdot
    \left(\frac{s}{s\sbs{min}}\right)^{3q\sbs{p}-5}
    \!\!\!\!\!\!
    \cdot \frac{r^{5/2}\d r}{\left(\mathsf{G} M_*\right)^{1/2}}
    \nonumber\\
    &\times& \frac{I}{f G(q\sbs{p},s)}.
    \label{eqtausig}
\end{eqnarray}

In order to replace the dependence on the initial cross
sectional area of objects, $\sigma\sbs{tot}$, with their initial total mass,
$M_0$, we need to derive both quantities from the initial size distribution in
Eq.~(\ref{eqDist0}).
The area is given by
\begin{equation}
  \sigma\sbs{tot} = n\sbs{max}(0) \cdot \frac{\pi s\sbs{max}^3}{3q\sbs{p}-5}
    \left[\left(\frac{s\sbs{max}}{s\sbs{min}} \right)^{3q\sbs{p}-5} - 1\right].
\end{equation}
Since it is dominated by $s\sbs{min}$ for $q\sbs{p} > 5/3$, we obtain
\begin{equation}
  \sigma\sbs{tot} = n\sbs{max}(0) \cdot \frac{\pi s\sbs{max}^3}{3q\sbs{p}-5}
    \left(\frac{s\sbs{max}}{s\sbs{min}} \right)^{3q\sbs{p}-5}.\label{eqSigma}
\end{equation}
The initial total disk mass is
\begin{equation}
  M_0 = n\sbs{max}(0) \cdot \frac{4\pi\rho
    s\sbs{max}^4}{3(6-3q\sbs{p})}
    \left[
      1 - \left(\frac{s\sbs{min}}{s\sbs{max}} \right)^{6-3q\sbs{p}}
    \right].
    \label{eqM0}
\end{equation}
For $q\sbs{p} < 2$, it is dominated by $s\sbs{max}$. However,
since a primordial slope $q\sbs{p} \geq 2$ is not unrealistic (see
Sect.~\ref{secParameters}) we refrain from using a further approximation.
Then, the area and the mass are related through
\begin{eqnarray}
  \sigma\sbs{tot} &=& M_0 \cdot
     \frac{3(2-q\sbs{p})}{4(q\sbs{p} - 5/3)}\cdot s\sbs{max}^{-1} \cdot
     \left(\frac{s\sbs{max}}{s\sbs{min}}\right)^{3q\sbs{p} - 5}
     \nonumber\\
      &\times&
    \left[
      1 - \left(\frac{s\sbs{min}}{s\sbs{max}} \right)^{6-3q\sbs{p}}
    \right]^{-1}.
  \label{eqM0Sigma}
\end{eqnarray}
Inserting Eq.~(\ref{eqM0Sigma}) into Eq.~(\ref{eqtausig}) results in
\begin{eqnarray}
  \tau(s) &=&
    \frac{16 \pi\rho}{3 M_0} \cdot
    \left(\frac{s}{s\sbs{max}}\right)^{3q\sbs{p}-5}
    \frac{s\sbs{max} r^{5/2} \d r}{\left(\mathsf{G} M_*\right)^{1/2}}
    \nonumber\\
    &\times&
    \frac{q\sbs{p}-5/3}{2-q\sbs{p}}
    \left[
      1 - \left(\frac{s\sbs{min}}{s\sbs{max}} \right)^{6-3q\sbs{p}}
    \right]
    \nonumber\\
    &\times&
    \frac{I}{f(e,I)G(q\sbs{p},s)},
    \label{eqtauM0}
\end{eqnarray}
which gives the collisional lifetime of an object with radius $s$.
Note that
\begin{equation}
  \frac{1}{2-q\sbs{p}}
  \left[
    1 - \left(\frac{s\sbs{min}}{s\sbs{max}} \right)^{6-3q\sbs{p}}
  \right]
  \longrightarrow
  3\:\ln \frac{s\sbs{max}}{s\sbs{min}}
\end{equation}
for $q\sbs{p} \rightarrow 2$.

If the mean impact velocities in the system are high enough to allow
planetesimals of radius $s$ to get disrupted in a collision, i.e.
$X\sbs{c}(s) \ll s\sbs{max}/s$, $G(q\sbs{p},s)$ reduces to
\begin{equation}
  G(q\sbs{p}, s) \approx \frac{q\sbs{p}-5/3}{q\sbs{p}-1} \cdot
  X\sbs{c}(s)^{3-3q\sbs{p}},\label{eqGRed}
\end{equation}
and $\tau(s)$ to
\begin{eqnarray}
  \tau(s) &=& \frac{16 \pi\rho}{3 M_0} \cdot
    \left(\frac{s}{s\sbs{max}}\right)^{3q\sbs{p}-5}
    \cdot r^2 \d r
    \cdot \left(\frac{r}{\mathsf{G} M_*}\right)^{q\sbs{p} - 1/2}
      \nonumber\\ &\times&
    \frac{q\sbs{p}-1}{2-q\sbs{p}}
    \left[
      1 - \left(\frac{s\sbs{min}}{s\sbs{max}} \right)^{6-3q\sbs{p}}
    \right]
    \cdot \left(2Q\sbs{D}^*\right)^{q\sbs{p}-1}
      \nonumber\\ &\times&
    \frac{I}{f(e,I)^{2q\sbs{p}-1}}
    \label{eqtaured}
\end{eqnarray}
Now, we take into account the dependence of $Q\sbs{D}^*$ on the
object size $s$, as was done by \citet{o'brien-greenberg-2003}.
If we are only interested in the gravity regime, $s > s\sbs{b}$,
Eq.~(\ref{eqQD}) is simplified to
\begin{equation}
  Q\sbs{D}^* (s) \approx Q\sbs{D,b}^*\cdot\left(\frac{s}{s\sbs{b}}\right)^{3b\sbs{g}},
\end{equation}
where $Q\sbs{D,b}^*$ is the critical specific energy at the
breaking radius, i.e. around the minimum of $Q\sbs{D}^*(s)$.
Assuming, further, that $I \propto e$,
we can write down the dependencies of the collisional lifetime,
\begin{equation}
  \tau(s) \propto \sigma\sbs{tot}^{-1} \cdot
      s^{3q\sbs{p}-5+3(q\sbs{p}-1)b\sbs{g}} \cdot
      r^{3/2+q\sbs{p}} \cdot \d r \cdot e^{-5/3}.\label{eqDependencies}
\end{equation}
\citet{o'brien-greenberg-2003} yield the same size dependence on $s$ in their
Eq.~(11).

To find $s\sbs{t}(t)$, the object size below which a steady state is reached,
we assume that the populations move from their primordial state to the
quasi-steady state instantaneously when the system age reaches their initial mean
collisional lifetime, $\tau(s\sbs{t}) = t$.
Inverting that, the resulting mass of objects in transition can be retrieved
as a function of system age. Keeping the assumption $X\sbs{c} \ll s\sbs{max}/s$,
the relation is
\begin{eqnarray}
  s\sbs{t} (t) &\propto&
    t^{1/(3q\sbs{p}-5+3(q\sbs{p}-1)b\sbs{g})}
\label{eqEquiMass}
\end{eqnarray}
for $t > \tau(s\sbs{b}) \equiv \tau\sbs{b}$.
This transitional size is also plotted in Fig.~\ref{figMassDistribution}.

\citet{pan-sari-2005} followed a similar approach in their study of the Kuiper-belt
size  distribution. Describing the propagation of the shock wave
through the target, they introduce a parameter $\beta$
that varies between $3/2$ (if all energy of a projectile goes to the shock wave)
and $3$ (if all its momentum does).
Their $\beta$ equals $1/b\sbs{g}$ in our nomenclature, and $b\sbs{g} = 0.5$
leads to $\beta = 2$.
Additionally, we have to replace their
slope $q_0$ with our $3q\sbs{p} - 2$.
Then, given their Eqs.~(6), (7), and $N_{>s} \propto s^{3-3q\sbs{p}}$,
we yield the same exponent as in our Eq.~(\ref{eqEquiMass}).
Note that what \citet{pan-sari-2005} call ``breaking radius'' is our
``transition radius'' $s\sbs{t}$, and 
their ``radius of equilibrium'' is our ``breaking radius'' $s\sbs{b}$.

\subsection{Evolution of Disk Mass}

Now, we derive the full expression for the time-dependent total disk mass.
Using the size distribution given by Eq.~(\ref{eqnm}),
$n\sbs{max}$ from Eq.~(\ref{DD}) and expressing
$n\sbs{max}(0)$ through $M_0$ with the aid of Eq.~(\ref{eqM0}),
we can perform the integration in Eq.~(\ref{eqTotalMass}).
Then, the resulting time-dependent disk mass is
\notetoeditor{Using ``frac'' instead of ``case'' is preferred but problematic
with respect to formula layout. Note that the text refers to individual lines
of the following equation.}
\begin{eqnarray}
  M\sbs{disk}(t) &=&
    \frac{M_0}{1+t/\tau\sbs{max}}
     \left[
     1 - \left(\frac{s\sbs{min}}{s\sbs{max}} \right)^{6-3q\sbs{p}}
    \right]^{-1}
    \nonumber\\
    &\times&
       \left[
            1
          - \left( \case{s\sbs{t}(t)}{s\sbs{b}} \right)^{6-3q\sbs{p}}
            \left( \case{s\sbs{b}}{s\sbs{max}} \right)^{6-3q\sbs{p}}
            \left( 1 - \case{2-q\sbs{p}}{2-q\sbs{g}} \right)
       \right. \nonumber\\
    &+&
         \left( \case{s\sbs{t}(t)}{s\sbs{b}} \right)^{3q\sbs{g} - 3q\sbs{p}}
         \left( \case{s\sbs{b}}{s\sbs{max}} \right)^{6-3q\sbs{p}}
         \left( \case{2-q\sbs{p}}{2-q\sbs{s}} -
                \case{2-q\sbs{p}}{2-q\sbs{g}}
         \right) \nonumber\\
    &-&
         \left( \case{s\sbs{t}(t)}{s\sbs{b}} \right)^{3q\sbs{g} - 3q\sbs{p}}
         \left( \case{s\sbs{b}}{s\sbs{max}} \right)^{3q\sbs{s} - 3q\sbs{p}}
         \left( \case{s\sbs{min}}{s\sbs{max}} \right)^{6-3q\sbs{s}}\nonumber\\
    && \left.
      \times \left( \case{2-q\sbs{p}}{2-q\sbs{s}} \right)
      \right]\label{eqMassContrib}
\end{eqnarray}
for $\tau\sbs{b} < t < \tau\sbs{max}$.
To make Eq.~(\ref{eqMassContrib}) valid for earlier phases,
i.e. for $t < \tau\sbs{b}$, $s\sbs{b}$ should be replaced by
$s\sbs{t}(t)$. The sizes involved are the maximum object size
$s\sbs{max}$, the transition size between the primordial and reprocessed
material $s\sbs{t}$, the breaking radius between the gravity and strength regime
$s\sbs{b}$. The lower limit in the size
distribution, $s\sbs{min}$, is crucial for the
dust emission and it is usually taken to be the radiation pressure blowout
limit. As long as $q\sbs{p} < 2$, it is fairly unimportant for the mass
budget. However, we are interested in $q\sbs{p} \geq 2$
as well. Therefore, we
can safely set $s\sbs{min} = 0$ only in the last line of
Eq.~(\ref{eqMassContrib}), where it enters through $s\sbs{min}/s\sbs{max}$
to the power of $6-3q\sbs{s}$, with $q\sbs{s}\approx 11/6 < 2$.

\begin{figure}[t!]
  \includegraphics{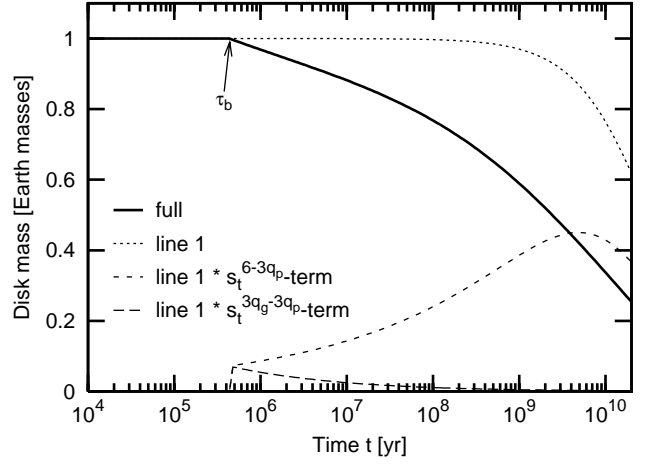}\\
  \caption{%
    The contributions of different terms in Eq.~(\ref{eqMassContrib})
    (dotted and dashed lines) and their total (solid line).
    \label{figMassContribution}}
\end{figure}

The relative importance of the terms in Eq. (\ref{eqMassContrib}) is
illustrated in Fig.~\ref{figMassContribution}.
A combination of the classic Dominik-Decin behavior in the first line of
Eq.~(\ref{eqMassContrib}) together with the second line is a reasonably accurate
approximation to $M\sbs{disk}(t)$ for most of the time.
With the aid of Eq.~(\ref{eqEquiMass}),
Eq. (\ref{eqMassContrib}) transforms to
\begin{eqnarray}
  M\sbs{disk}(t) &\approx&
    \frac{M_0}{1+t/\tau\sbs{max}}
     \left[
     1 - \left(\frac{s\sbs{min}}{s\sbs{max}} \right)^{6-3q\sbs{p}}
    \right]^{-1}
    \nonumber\\
  &\times&
  \left[
       1 - \left( \frac{s\sbs{b}}{s\sbs{max}} \right)^{6-3q\sbs{p}} \cdot
           \left( \frac{t}{\tau\sbs{b}} \right)^
              {\frac{2-q\sbs{p}}{q\sbs{p}-5/3+(q\sbs{p}-1)b\sbs{g}}}
           \right.\nonumber\\
       && \left.\times
           \left( 1 - \frac{2-q\sbs{p}}{2-q\sbs{g}} \right)
  \right]\label{eqEquiMassShort}
\end{eqnarray}
for $\tau\sbs{b} < t < \tau\sbs{max}$.
At $t \ll \tau\sbs{max}$, and assuming $q\sbs{p} = 1.87$,
a further approximation is
\begin{equation}
  M\sbs{disk}(t) \approx M_0
  \left(
       1 - {\rm const} \cdot t^{0.2}
  \right).
\end{equation}

\begin{figure}[t!]
  \includegraphics{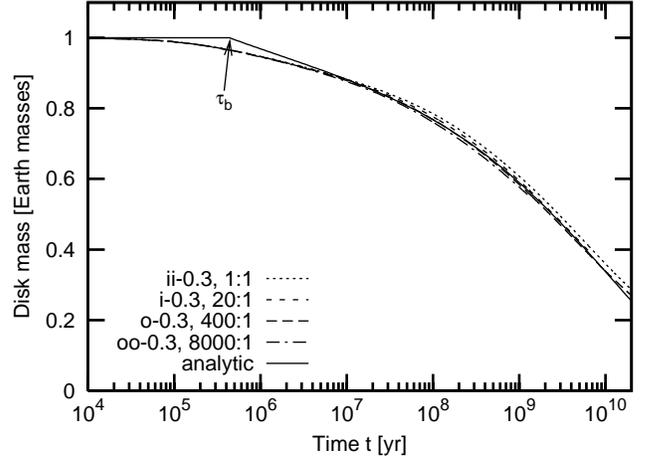}\\
  \caption{Evolution of total masses with (scaled) time,
    obtained in four numerical runs and with the analytic model.
  \label{figTotalMassAnaNum}}
\end{figure}

The evolution of the disk mass, both from the numerical runs and
from the analytic solution (\ref{eqMassContrib}),
is plotted in Fig.~\ref{figTotalMassAnaNum}, showing a good agreement
between analytics and numerics.
A deviation is only seen around $t = \tau\sbs{b}$ where the transition
from primordial to reprocessed state sets on for gravity-dominated objects.
The reason is that, to ease the analytic treatment, we neglect the smooth
natural transition from material strength to self-gravity given by
Eq.~(\ref{eqQD}) and assume a sharp break between the two power laws instead.

\subsection{Evolution of Disk Mass at Latest Stages}

As soon as the age of the system has reached the collisional lifetime of the
largest bodies, i.e. at $t > \tau\sbs{max}$, the solids of all sizes in the disk
reach quasi-steady state, and the change in total mass will be dominated
by $1/t$.
At this latest phase,
the projectiles that
can destroy objects of size $s\sbs{max}$ no longer follow a
size distribution with the primordial slope, $2-3q\sbs{p}$.
Instead, they have the slope of a
collisional cascade under gravity regime, $2-3q\sbs{g}$. The slightly
longer collisional lifetime can neither be expressed through
Eq.~(\ref{eqtc}) that uses the \emph{initial} cross section
$\sigma\sbs{tot}$
nor through Eq.~(\ref{eqtauM0}) that contains the initial disk mass $M_0$ and
slope $q\sbs{p}$.
The correct way to evaluate $\tau\sbs{max}$ is to use the initial number density
of biggest objects, $n\sbs{max}(0)$, and the slope $q\sbs{g}$.
Expressing $\sigma\sbs{tot}$ in Eq.~(\ref{eqtc}) through $n\sbs{max}$ with
the help of Eq.~(\ref{eqSigma}) and replacing then $q\sbs{p}$ with $q\sbs{g}$,
we obtain
\begin{eqnarray}
  \tau\sbs{max} &=&
        \frac{12q\sbs{g}-20}{n\sbs{max}(0)\cdot s\sbs{max}^3} \cdot
        \frac{r^{5/2} \d r}{\left(\mathsf{G} M_*\right)^{1/2}}
\nonumber\\
&\times&
        \frac{I}{f(e,I)G(q\sbs{g}, s\sbs{max})}.
\end{eqnarray}
Expressing now $n\sbs{max}(0)$ through $M_0$ by virtue of Eq.~(\ref{eqM0})
yields
\begin{eqnarray}
  \tau\sbs{max} &=&
    \frac{16 \pi\rho}{3 M_0} \cdot
s\sbs{max} \cdot
    \frac{r^{5/2} \d r}{\left(\mathsf{G} M_*\right)^{1/2}}
  \nonumber\\
  &\times&
    \frac{q\sbs{g}-5/3}{2-q\sbs{p}}
    \left[
      1 - \left(\frac{s\sbs{min}}{s\sbs{max}} \right)^{6-3q\sbs{p}}
    \right]^{-1}
  \nonumber\\
  &\times&
    \frac{I}{f(e,I)G(q\sbs{g}, s\sbs{max})}
    \label{eqtcEx},
\end{eqnarray}
where both slopes, $q\sbs{p}$ and $q\sbs{g}$, appear
(cf. Eqs.~\ref{eqtc} and \ref{eqtauM0}).

\subsection{Evolution of Mass in Dynamically ``Cold'' Disks}

All the treatment above applies to planetesimal belts where relative velocities 
are high enough for the biggest objects to be destroyed by
mutual collisions. This might not be the case in dynamically ``cold'' disks
with low eccentricities and inclinations and/or very far from the star.

Consider again the lifetime of objects $\tau(s)$.
As $s$ increases, $X\sbs{c}(s)$ (Eq.~\ref{eqXcFull})
increases too and at a certain point reaches $s\sbs{max}/s$.
At this point, $G$ (Eq.~\ref{eqGFull}) becomes zero and $\tau(s)$  (Eq.~\ref{eqtauM0})
goes to infinity.
This means that, for a given impact velocity, objects above a certain
critical size cannot be disrupted anymore.
In systems with low relative velocities, that critical size may happen
to be smaller than $s\sbs{max}$. This will affect the mass evolution.
Specifically, when $s\sbs{t}$ reaches that critical size, the overall
mass decay ceases.

\begin{figure}[t!]
  \includegraphics{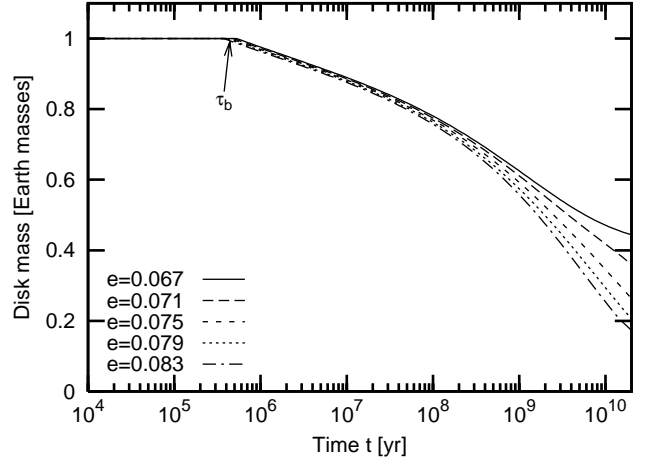}\\
  \caption{
    Influence of the effective eccentricity assumed in the analytic model
    for a disk of 1~$M_\oplus$ at $r=10$~AU with a radial extent
    $\d r=7.5$~AU. The $I=e/2$ relation between eccentricity and inclination
    is assumed.
  \label{figQcomp}}
\end{figure}

To illustrate such effects,
Fig.~\ref{figQcomp} shows the influence of the effective
$e$ and $I$ on the evolution of the total mass of a disk of initially
1~$M_\oplus$ at an effective distance of $10$~AU, calculated with our
analytic model.
For colder disks, the curves start to flatten. This happens because the
largest planetesimals (that dominate the total mass) stay intact, which
slows down the mass loss.

\subsection{Evolution of Dust Mass}

\begin{figure}[t!]
  \includegraphics{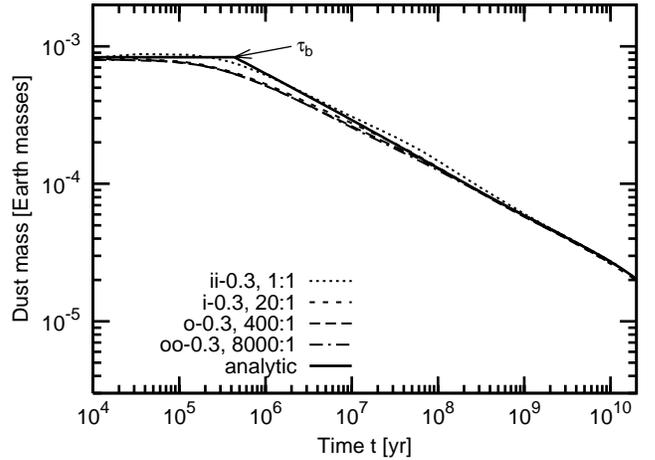}\\
  \caption{%
    Similar to Fig. \ref{figTotalMassAnaNum} but for dust masses, i.e.
    masses in particles with radii below 1~mm.
  \label{figDustMassAnaNum}}
\end{figure}
The dust mass can be evaluated in a similar way as the disk mass.
We use now Eqs.~(\ref{eqnm}), (\ref{eqDustMass}), (\ref{DD}), 
(\ref{eqM0}), and (\ref{eqEquiMass}).
Neglecting the minimum mass $s\sbs{min}$ only when
it enters the formula through $s\sbs{min}/s\sbs{max}$, we obtain
\begin{eqnarray}
  M\sbs{dust} (t) = \frac{M_0}{1+t/\tau\sbs{max}}\cdot
        \left(\frac{t}{\tau\sbs{b}}\right)^
        {\frac{q\sbs{g}-q\sbs{p}}{q\sbs{p}-5/3+(q\sbs{p}-1)b\sbs{g}}}\cdot
        \frac{2-q\sbs{p}}{2-q\sbs{s}}
      \nonumber\\
      \times
      \left(\frac{s\sbs{b}}{s\sbs{max}}\right)^{2-q\sbs{p}}
      \left[\left(\frac{s\sbs{d}}{s\sbs{b}}\right)^{2-q\sbs{s}} -
            \left(\frac{s\sbs{min}}{s\sbs{b}}\right)^{2-q\sbs{s}}
         \right]^{-1}
        \label{eqDustMassEx}
\end{eqnarray}
for $\tau\sbs{b} < t < \tau\sbs{max}$.
Before that, i.e. at $t < \tau\sbs{b}$,
we have $q\sbs{s}$ and $b\sbs{s}$ instead of $q\sbs{g}$ and $b\sbs{g}$,
respectively. If the assumed primordial slope, $q\sbs{p}$, equals the
steady-state slope in the strength regime, $q\sbs{s}$,
the dust mass stays constant, which is the case for the first part of the
numerical integration.
However, as soon as the transitional zone reaches objects large enough to be
influenced by self-gravity,
Eq.~(\ref{eqDustMassEx}) starts to work.
It shows that the evolution of dust mass depends most strongly
on the difference between $q\sbs{p}$ and $q\sbs{g}$.
The dust mass decay, obtained both from the numerical runs and
analytic solution (\ref{eqDustMassEx}),
are shown in Fig.~\ref{figDustMassAnaNum}. For $t > \tau\sbs{b}$,
we roughly have $M\sbs{dust} \propto t^\xi$ with $\xi \approx -0.3$.

We finally note that Eq.~(\ref{eqDustMassEx}) is valid as long as the
collisional lifetime
of the largest planetesimals is longer than the age of the system.
When $t > \tau\sbs{max}$, $t/\tau\sbs{b}$ in that equation must be replaced by 
$\tau\sbs{max}/\tau\sbs{b}$.

\subsection{The Model Parameters}
\label{secParameters}

Our analytic model contains several parameters that either
differ from similar parameters in the numerical model (such as $e$) or are absent there
(such as $q\sbs{s}$ and $q\sbs{g}$).
To use the analytic model, we have to specify them.
We now describe how this can be done, explaining, in particular, 
the choice of parameters used to plot analytic curves in
Figs.~\ref{figMassContribution}--\ref{figDustMassAnaNum}.

Two important free parameters of the analytic model are $q\sbs{s}$ and $q\sbs{g}$.
We use the work of \citet{o'brien-greenberg-2003} who
found the slope of the size distribution in a system in a collisional steady
state. With the dependence of the critical specific energy on the object
size given in Eq.~(\ref{eqQD}), they give a power-law index
\begin{eqnarray}
  q = \frac{11/6+b}{1+b} \label{eqOBrienq}
\end{eqnarray}
in their Eq.~(24). With $b=b\sbs{s}=-0.1$ for the strength regime we have
$q=q\sbs{s}=1.877$.
Similarly, with $b=b\sbs{g}=0.5$ for the gravity regime,
Eq.~(\ref{eqOBrienq}) can be used to derive
$q\sbs{g} \approx 5/3$.
It is these values that we used in 
Eq.~(\ref{eqMassContrib}) to produce Figs.~\ref{figMassContribution}--\ref{figQcomp}
and in Eq.~(\ref{eqDustMassEx}) to plot Fig.~\ref{figDustMassAnaNum}.

\begin{figure}[t!]
  \includegraphics{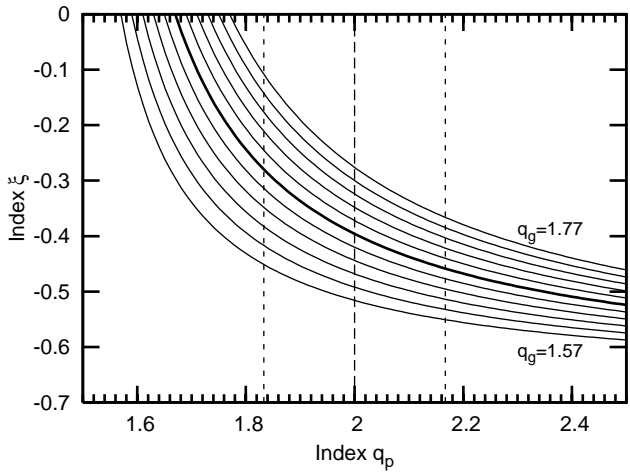}\\
  \caption{Index $\xi$ of the power-law evolution of the dust mass,
  $M\sbs{d}\propto t^\xi$. The horizontal axis gives the dependence on the
  slope of the primordial mass distribution, $q\sbs{p}$, for values from
  $q\sbs{g} = 1.57$ (bottom) to $q\sbs{g} = 1.77$ (top) for the slope in the
  gravity regime. The bold line is for $q\sbs{g} = 1.67 \approx 5/3$.
  Vertical lines indicate the mean value and error estimates for
  $q\sbs{p}$ from \citet{trujillo-et-al-2001}.
  \label{figDustEvoSlope}}
\end{figure}

In contrast to $q\sbs{s}$ and $q\sbs{g}$, the primordial slope, $q\sbs{p}$,
is a free parameter not only in the analytic model, but also in the numerical one.
As stated in Sect. ~\ref{secCommons}, in all ``nominal'' runs we assumed
$q\sbs{p} = 1.87$, which corresponds to $p\sbs{p} = 3q\sbs{p} -2 = 3.61$ in the size scaling.
In principle, $q\sbs{p}$ describes the mass
distribution at the onset of the collisional grinding
of the disk and, therefore, represents a link to the planetesimal formation
process. The outcome of the agglomeration phase is the input to the phase of
disruptive collisions. 
The Kuiper belt
is the only source for observational constraints to this parameter so far, and
recent surveys suggest a value of $p\sbs{p} = 4.0 \pm 0.5$
\citep[e.g.,][]{trujillo-et-al-2001, bernstein-et-al-2004} or $q\sbs{p} =
2.00 \pm 0.17$. Simulations by \citet{kenyon-bromley-2004c} yield
$p\sbs{p}=4.0$--$4.5$ or $q\sbs{p}=2.00$--$2.17$. According to
Eq.~(\ref{eqDustMassEx}), where we have $M\sbs{dust} \propto t^\xi$, and
together with $q\sbs{g} \approx 1.67$, this would
change the dust mass evolution from $M\sbs{dust}\propto t^{-0.32}$ for $q\sbs{p}
= 1.87$ to $M\sbs{dust}\propto t^{-0.40}$ for $q\sbs{p} = 2.00$.
Fig.~\ref{figDustEvoSlope} shows the rather moderate dependence of the index
$\xi$ on the two mass distribution slopes, $q\sbs{g}$ and $q\sbs{p}$.

While the dust size limit, $s\sbs{d}$, has little influence on the
mass budget, the breaking size, $s\sbs{b}$, the maximum size, $s\sbs{max}$,
and the ratio of the two are relevant to the evolution as they define the
lifetime of the largest bodies $\tau\sbs{max}$ relative to $\tau\sbs{b}$.
What is more, the ratio $s\sbs{b}/s\sbs{max}$ determines the rate of the mass
decay in Eq.~(\ref{eqEquiMassShort}).
From Sect.~\ref{secQD} we know the location of the breaking radius to be
$316$~m for the material properties assumed, and the upper size limit
of all the runs was set to $s\sbs{max}=74$~km.

Another parameter in the analytic model is the collisional lifetime of objects of breaking radius,
$\tau\sbs{b} = \tau(m\sbs{b})$.
Eq.~(\ref{eqtauM0}) expresses it through other
parameters critical for the efficiency of 
collisions: the radial distance to the star $r$, the disk radial extension
$\d r$, and the effective eccentricity $e$
and inclination $I$.
We choose to fix both the effective distance and the disk
extension to be  $r = 4/3 \d r = 10$~AU when reproducing analytically the
results of the ii-0.3 run, 20~AU for i-0.3, 40~AU for o-0.3, and 80~AU for oo-0.3.
Further, the inclination can be coupled to eccentricity
by assuming the equilibrium condition $I = e/2$.
Thus, only $e$ remains as a free parameter.
The best fit to, e.g., the ii-0.3 run is achieved if we assume $e\approx
0.075$ in the analytic model, which is approximately one quarter of
$e\sbs{max}=0.3$.
With these choices, we find $\tau(s\sbs{b}) \approx 4\times 10^5$~years.

Alternatively, $\tau\sbs{b}$
can be directly retrieved from the break
in the evolution of the dust mass (see Fig.~\ref{figDustMassAnaNum}).
This method  gives $\tau(s\sbs{b}) \approx 5\times 10^5$~years, 
which is approximately $4/3$  times the value calculated with Eq.~(\ref{eqtauM0}).
This discrepancy is probably a result of the
particle-in-a-box assumptions made by \citet{wyatt-et-al-2007a} in derivation of
Eq.~(\ref{eqtc}).
We prefer this empirical scaling and thus applied the factor of $4/3$ to all analytically
estimated timescales in this paper.

\section{EVOLUTION OF DISK LUMINOSITY}
\label{secEvolLuminosity}

\subsection{Fractional Luminosity for a Given Age}
\label{sec_fmax}

Following \citet{wyatt-et-al-2007a}, we define the
fractional luminosity of dust as
\begin{equation}
  f\sbs{d} \equiv \sigma\sbs{tot} / (4 \pi r ^2) ,
  \label{f}
\end{equation}
which assumes that dust grains are black bodies, absorbing
and re-emitting all the radiation they intercept.
\citet[][their Eq.~20]{wyatt-et-al-2007a}
found that there is a maximum possible
fractional luminosity $f\sbs{max}$ for a given age, whose
value is independent of the initial disk mass, but depends on
other model parameters such as
the distance $r$ of the disk center from the star,
its width $\d r$,
size of the largest planetesimals $D\sbs{c}$,
critical fragmentation energy $Q\sbs{D}^*$,
orbital eccentricity of planetesimals $e$
(with their inclination being $I = e/2$),
as well as the stellar mass $M_*$ and luminosity $L_*$.

\begin{figure*}[t!]
  \centering
  \includegraphics[width=0.5\linewidth]{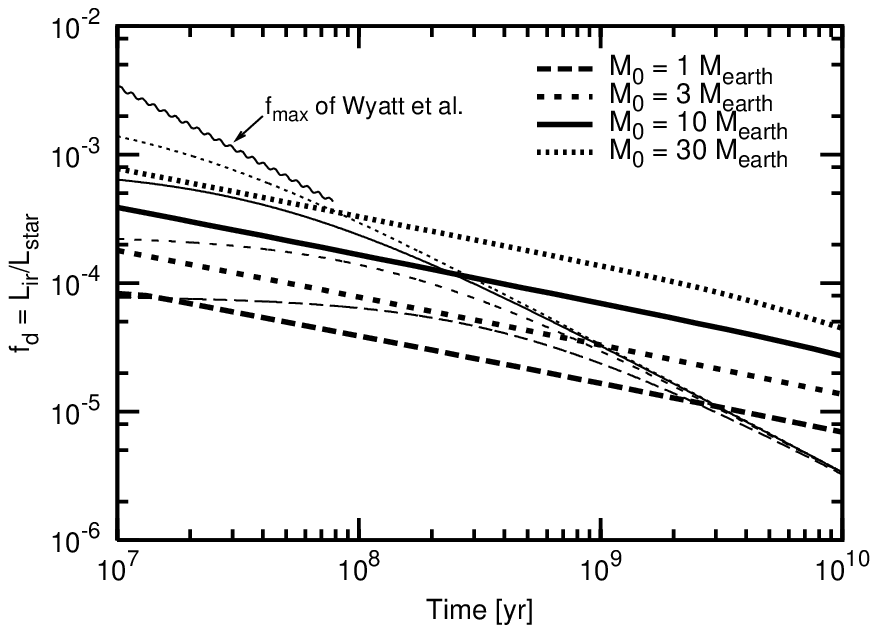}%
  \includegraphics[width=0.5\linewidth]{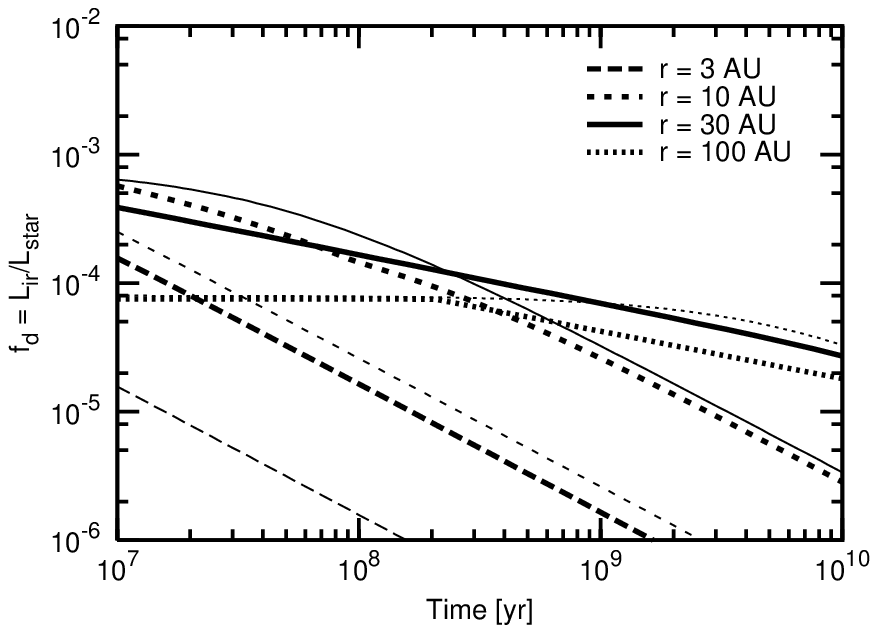}\\
  \includegraphics[width=0.5\linewidth]{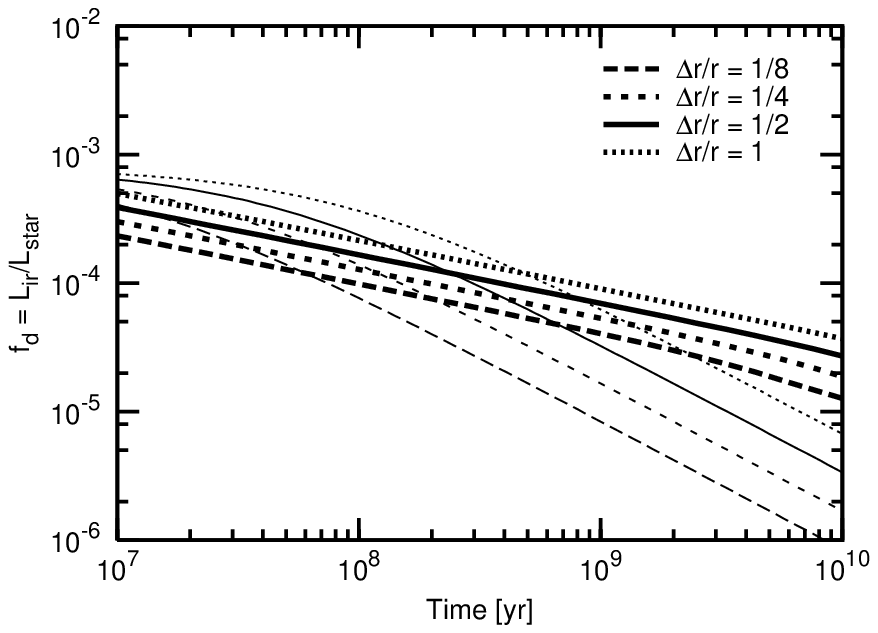}%
  \includegraphics[width=0.5\linewidth]{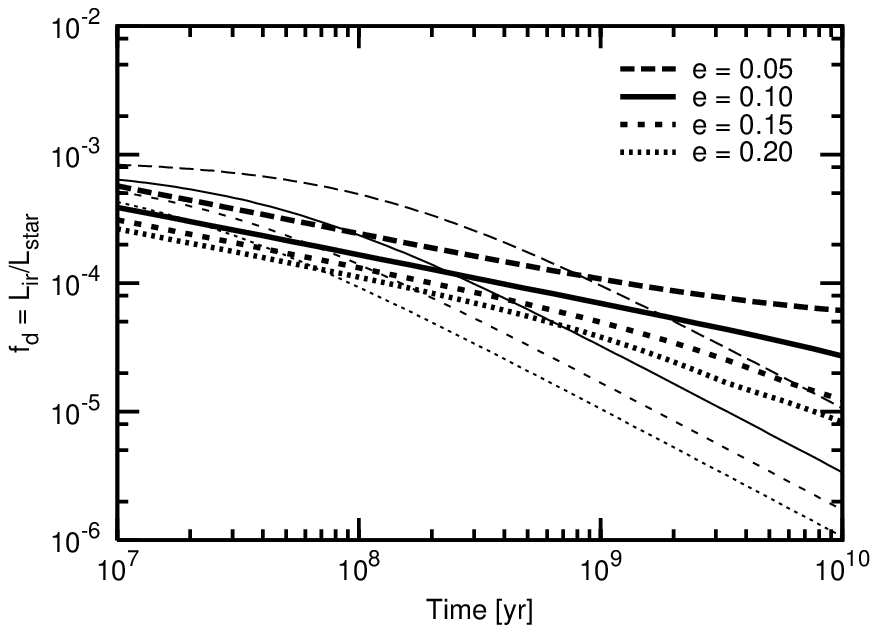}
  \notetoeditor{The "width" parameter should be obsolete in the final two-column version.}
  \caption{
  Fractional luminosity of dust around a solar-like star as a function of age.
  Thick lines: our analytic model; thin lines: $f\sbs{d}$ of \citet{wyatt-et-al-2007a}.
  Different panels demonstrate dependence on different parameters:
  $M\sbs{disk}$ \textsl{(top left)},
  $r$ \textsl{(top right)},
  $\d r/r$ \textsl{(bottom left)},
  and $e$ \textsl{(bottom right)}.
  A standard case with
  $M_* = L_* = 1$, $M\sbs{disk} = 10 M_\oplus$,
  $r = 30$~AU, $\d r/r = 1/2$, and $e = 0.10$ is shown
  with solid lines (common in all panels).
  \label{fig_fmax}}
\end{figure*}

We now wish to explore $f_d(t)$ and check whether it has an upper limit
in the framework of our analytic model.
To this end, we used Eq.~(\ref{f}) and calculated $\sigma\sbs{tot}$
with the aid of our Eq.~(\ref{eqDustMassEx}) for the dust mass.
We assumed a solar-type star with $M_* = L_* = 1$ and probed
disks with $M\sbs{disk} = 1, 3, 10$, and $30 M_\oplus$;
$r = 3, 10, 30$, and $100$~AU;
$\d r/r = 1/8, 1/4, 1/2$, and $1$;
$e = 0.05, 0.10, 0.15$, and $0.20$.
The results are presented in Fig.~\ref{fig_fmax} (thick lines).
As a standard case, we adopted
$M\sbs{disk} = 10 M_\oplus$,
$r = 30$~AU, $\d r/r = 1/2$, and $e = 0.10$.
It is shown with a thick solid curve in each of the panels.

In the same Fig.~\ref{fig_fmax}, we have overplotted with thin lines
the dust luminosity $f\sbs{d}$ computed with Eqs. (14), (19), and (20) of
\citet{wyatt-et-al-2007a}, for comparison.
In that calculation, we assumed $Q\sbs{D}^*=300$~J/kg (constant in their model),
$D\sbs{c} = 60\km$, and the same values
of those parameters that are common in their and our model
($M_*$, $L_*$, $r$, $\d r/r$, and $e$).

Analysis of Fig.~\ref{fig_fmax} allows us to make a number of conclusions.
First, as expected, our model yields more gently sloping curves
than that by Wyatt et al.
As discussed above, the $1/t$ law will be asymptotically reached in our model,
too, but this does rarely happen at ages $t < 10$~Gyr. Only the first signs
of the curves' steepening appear at Gyr ages, and that only for the cases when
the collisional evolution is faster (higher masses, closer-in or more confined
dust rings, higher eccentricities).
As a consequence of the slope difference between the two models,
our model places more stringent upper limits of $f\sbs{d}$ at earlier ages,
and conversely, it allows the Gyr-old systems to have a somewhat higher $f\sbs{d}$
than the model by Wyatt et al. does.

Next, the dependence of $f\sbs{max}$ on the initial disk mass, which cancels out in their
model, is retained in our nominal runs (the top left panel). In fact, the maximum 
possible $f\sbs{d}$ is then determined by the maximum initial disk mass that still appears
physically plausible in the framework of theories of planetesimal accretion and planet
formation.

Another point to mention is that, whereas the dependence on the disk width (bottom right)
and planetesimal eccentricities is relatively weak and monotonic, the dependence on
the disk location (top right) is rather strong and more intricate.
That the dependence is strong is the consequence of Eq.~(\ref{eqScalingR}) that predicts the
timescales to very sensitively depend on the distance from the star, and of Eq.~(\ref{f})
that contains a ``dilution factor'' $r^2$. At the beginning of the evolution, the innermost ring
is always the brightest because the dilution factor $r^2$ in  Eq.~(\ref{f}) is the smallest.
At the end of the evolution, the opposite is true: the outermost ring will become the brightest,
because its collisional evolution is the slowest and it retains more mass than
inner disks.
Therefore, all four curves intersect each other at a certain point;
the 30 and 100~AU curve do that after 10~Gyr, i.e. outside that right edge of the plot.
After that, all the curves go parallel to each other in the ``Dominik-Decin regime'',
following a $1/t$ law. Note that inner rings reach
the $1/t$ regime more quickly: already at 10~AU it is established in around 100~Myr for an
initial mass of 10~$M_\oplus$.

Although the existence of a ``maximum fractional luminosity for a given age'', as suggested by
\citet{wyatt-et-al-2007a}, no longer holds in our model as a robust mathematical
statement, in practice our model still suggests that $f\sbs{d}(t)$ cannot exceed a certain
limit, unless the model parameters take extreme values, incompatible with our understanding
of the planetesimal disks. For instance, we do expect $f\sbs{d} < 10^{-4}$ at $t=10$~Gyr,
provided that the initial disk did not contain more than 30 earth masses of solids and that
the mean orbital eccentricity of planetesimals is not lower than 0.1 (corresponding to
the mean inclination larger than $3^\circ$).
Therefore, plots such as Fig.~\ref{fig_fmax} can be used to check whether or not
$f\sbs{d}$ observed for a certain system with a known age is compatible with a ``smooth'',
unperturbed collisional evolutionary scenario. In case it is not, it will be an indication
that other mechanisms (delayed stirring, recent giant break-ups, non-collisional
dust production etc.) should be thought of to explain the observations.

\subsection{24 and 70 Micron Fluxes from Partial Rings}
\label{secLuminosity}
In order to produce directly observable quantities from the derived dust masses,
we now concentrate on dust luminosities at particular infrared wavelengths.
We calculated the dust temperature and the thermal emission integrated over the whole
disk with a more accurate, yet sufficiently simple, model, assuming that the
absorption/emission efficiency
is constant up to wavelengths of $2\pi$ times the size of the particles, $s$,
and proportional to $s^{-1}$ beyond that \citep{backman-paresce-1993}.
Then we computed the spectral flux densities
of dust emission $F\sbs{d}$ and of the stellar radiation $F_*$ at a certain wavelength,
as well as their ratio $F\sbs{d}/F_*$.
As the size distribution in the dust regime quickly reaches its steady state,
the luminosity $F\sbs{d}$ is directly proportional to the dust mass. Therefore, the same
initial constancy and subsequent $t^\xi$ decay with $\xi=-0.3 \ldots -0.4$ apply.

\begin{figure}[t!]
  \includegraphics{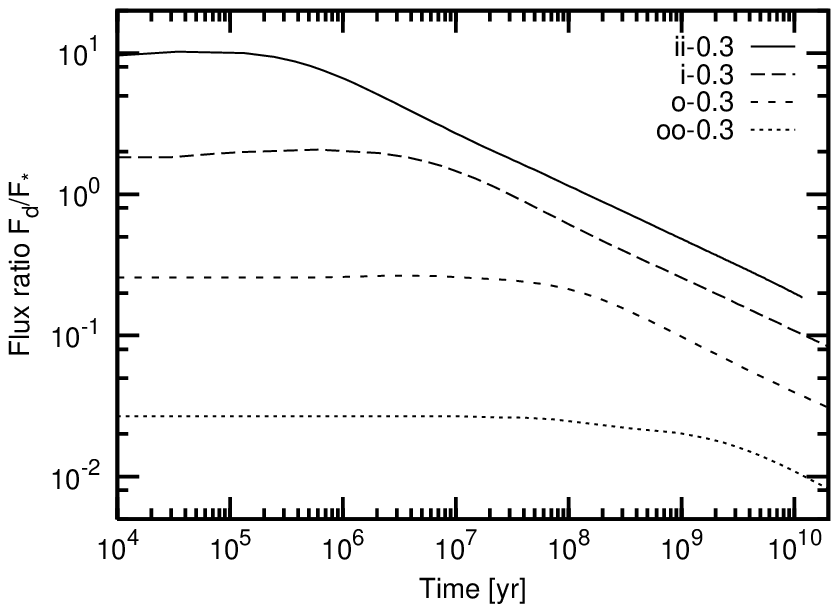}\\
  \includegraphics{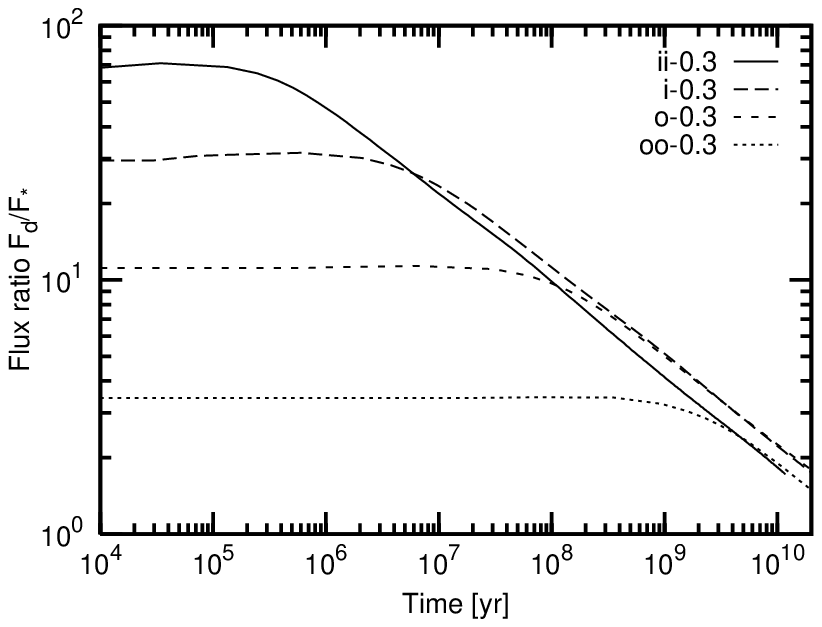}\\
  \caption{Flux ratio versus time for \textsl{(top)} 24~\textmu{m} and
  \textsl{(bottom)} 70~\textmu{m}.
  \label{fig2470}}
\end{figure}

Fig.~\ref{fig2470} shows the evolution of the excess
emission at the Spitzer/MIPS wavelengths 24 and 70~\textmu{m},
obtained from the four nominal runs.
Since all disks have the same initial total mass ($1$~$M_\oplus$), the disks
closer to the star are brighter and start to decay earlier.
The difference between the excesses at 24 and 70~\textmu{m}, a measure of the
disks' effective temperature, is varying with radial distance as well. Thus, the
convergence of just the 70~\textmu{m} fluxes at later times is only
coincidental. It is a result of the radial dependence of temperature
and the collisional timescale.

\subsection{Fluxes from Extended Disks}

Since resolved debris disks suggest that the parent body
reservoir in the disks is usually confined to a toroidal region
(a planetesimal belt), or is made up of several such tori,
it seems appropriate to simply combine individual rings without taking
into account possible interactions between particles that belong to different rings.
Thus, we summed up the fluxes from the four main runs. Different
radial distributions in the whole disk were simulated by ``weighting''
the individual rings:
\begin{equation}
  F\sbs{d} = \sum_{j=1}^{4}\limits F\sbs{d,j} (r_j/r_0)^{\gamma},
  \label{F_lambda}
\end{equation}
where $r_j$ are the central distances of the rings and
values of $0$, $1$, $2$, and $3$ were used for the slope $\gamma$. As the
reference runs were made for rings of one earth mass each with volumes
proportional to $r^3$, the corresponding volume density in the extended disk is
proportional to $r^{\gamma - 3}$, while the pole-on surface density and
normal geometrical optical depth follow $\propto r^{\gamma - 2}$.
The distance $r_0$ normalizes the total mass to $1 M_\oplus$.
Therefore, by changing the slope, the mass is only shifted between inner and outer regions.

\begin{figure}[t!]
  \includegraphics{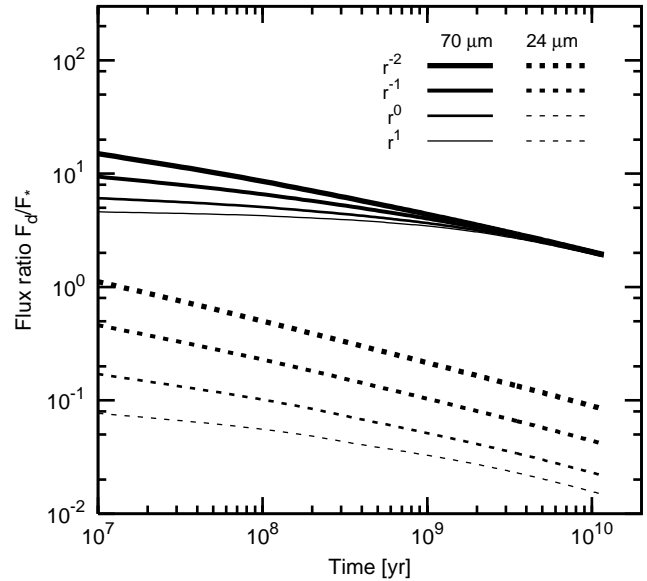}\\
  \caption{Time evolution of the infrared excess of extended disks
   with different initial radial distributions
   (labels indicate the radial slope of the  surface mass density;
   the thicker lines, the flatter the profiles)
   at  24 \textmu{m} (dashed lines) and 70 \textmu{m} (solid lines).
  The total mass is $1 M_\oplus$ in each case. 
  \label{figSummedFluxes}}
\end{figure}

In Fig. \ref{figSummedFluxes} the effect on the 24 and 70~\textmu{m} fluxes is
shown. If the weights are assigned in favor of more distant debris rings, the
resulting fluxes are naturally reduced. The same is true for the speed of
the decay because the timescales get longer. The evolution of the fluxes at the
two Spitzer/MIPS wavelengths 24 and 70~\textmu{m} differs significantly.
At 24~\textmu{m} the decay starts earlier and reaches its maximum speed earlier
because shorter-lived inner regions make the main contribution.

The models contain a sufficient number of parameters, variation of which
would  affect the curves in Fig. \ref{figSummedFluxes} in different ways.
As stated earlier, varying the total mass changes the timescale according to
$\tau \propto M\sbs{disk}^{-1}$. Hence,
the curves can be shifted along the lines of equal $t \cdot M\sbs{disk}$,
i.e. along the top left -- bottom right diagonal.
As seen from Fig.~\ref{figSummedFluxes}, variation
of the radial distribution changes both the absolute level and the tilt of the curves.
Besides, it affects the disk colors, i.e. the separation of the
24 and 70~\textmu{m} curves in Fig. \ref{figSummedFluxes}.
In addition, the dynamical timescales, and therefore the tilt of the curves,
are affected by eccentricities and inclinations of the parent bodies
that may reflect the presence of planetary perturbers in the disk
(see Sect.~\ref{secEcc}).
Altogether, these degrees of freedom would allow one to reproduce a broad set
of observational data.

\section{COMPARISON WITH OBSERVATIONAL DATA}
\label{secObservations}

\subsection{Spitzer Data}

The advent of the Spitzer Space Observatory has brought a tremendous 
increase in the number of main-sequence stars surveyed for the existence 
of cold dust emission (see \citet{werner-et-al-2006} for a recent compilation).

The wealth of data from these debris disk surveys allows us to confront 
our models with actual observations. To this end, we searched the literature 
for published flux ratios at 24 and/or 70 \textmu{m} (two of the three MIPS
bands) around G-type main-sequence stars. 
To qualify as a main-sequence star we applied a lower limit to the stellar age
of 10~Myr. Sources with stellar age estimates younger than this are likely
stars with gas-dominated, protoplanetary disks; these were not taken into
account.

The bulk of the data taken in the framework of the Legacy program ``Formation 
and Evolution of Planetary Systems'' (FEPS)
\citep{meyer-et-al-2004,meyer-et-al-2006} is
public since December 2006. The FEPS archive contains images, spectra,
photometry tables and Kurucz photosphere models and is available at
\url{http://\linebreak[0]data.spitzer.caltech.edu/\linebreak[0]%
popular/\linebreak[0]feps/\linebreak[0]20061223\_%
enhanced\_v1/}.
Age estimates have been published for 46 FEPS G stars
\citep{kim-et-al-2005,stauffer-et-al-2005,silverstone-et-al-2006}.

The large Guaranteed Time Observer (GTO) survey of FGK stars contains another 
64 stars, where ages are available
\citep{beichman-et-al-2005,beichman-et-al-2006b,bryden-et-al-2006}. Data for ten
more G stars are listed in \citet{chen-et-al-2005a,chen-et-al-2005b}.
In total, 120 G-type main-sequence stars with flux ratios at 24 and/or 70 $\mu$m have
been compiled from the literature for comparison with model flux ratios. 

\subsection{Population Synthesis}

Based on the analytic prescription presented in Sect.~\ref{secAnalytics} and
motivated by the \citet{wyatt-et-al-2007b} work, we now
build a synthetic set of debris disks around G2 stars. We
generate a set of ring-like disks of width $\d r$
located at distances $r \in [r\sbs{min},r\sbs{max}]$,
with masses $M\sbs{disk}\in [M\sbs{min}, M\sbs{max}]$,
and ages between 10~Myr and 10~Gyr.
The probability to have a disk of initial mass $M_0$ at radius $r$ was assumed
to follow $M_0^{-1} r^{-0.8}$,
where $M_0^{-1}$ corresponds to a log-normal distribution of initial disk
masses and the $r^{-0.8}$ dependence was proposed by \citet{wyatt-et-al-2007b}.
As described in Sect.~\ref{secLuminosity}, the temperatures and the resulting
thermal fluxes are calculated using the modified black-body formulas
by \citet{backman-paresce-1993}, assuming the emitting grains
to have $s=1$~\textmu{m},
in agreement with the size distribution shown in Fig.~\ref{figMassDistribution}.
The other parameters are taken to be: $q\sbs{p} = 2.00$, $q\sbs{g}=1.67$,
$q\sbs{s}=1.877$, $\d r/r=0.5$, $2I = e = 0.15$, $Q\sbs{D}^*(1\:\mathrm{m}) =
Q\sbs{D}^*(1\:\mathrm{km}) = 5\times 10^6$~erg/g, $b\sbs{d}=-0.12$,
$b\sbs{g}=0.47$,
roughly corresponding to basalt in \citet{benz-asphaug-1999}.
Due to the small observational sample, our aim was not to perform a multi-parameter 
fit to the observations, but rather to cover the range of observed flux densities, which is
defined by the limits of the distributions, not by their slopes.

\begin{figure}[t!]
  \includegraphics{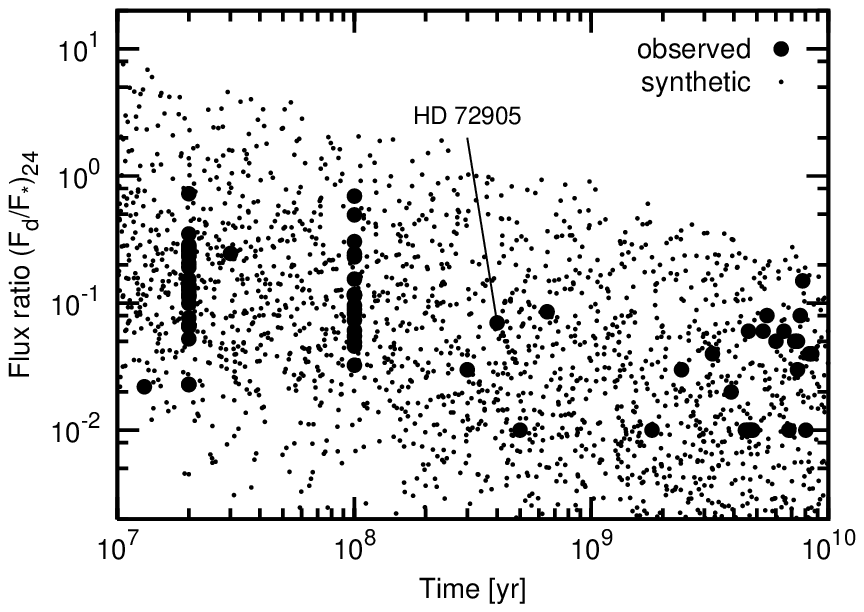}\\
  \includegraphics{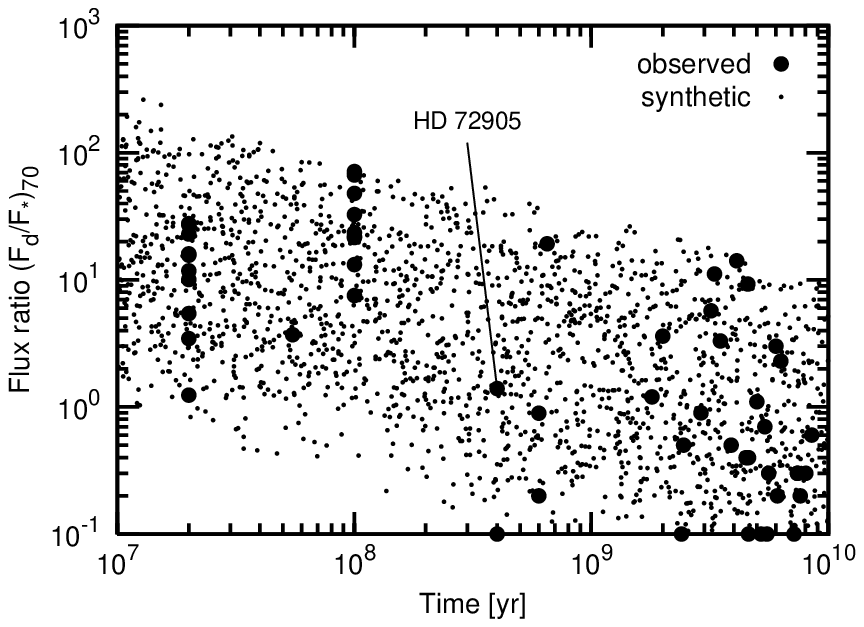}\\
  \caption{Flux ratios versus time for 24 \textmu{m} \textsl{(top)}
  and  70 \textmu{m} \textsl{(bottom)}.
  The synthesized population (\textsl{small dots}) is compared to the observed
  one (\textsl{big dots}).
  Individually labeled is the possibly transient system HD 72905, see text.
  \label{fig2470obs}}
\end{figure}

Varying disk locations and masses easily reproduces the observed
distribution of fluxes at 24~\textmu{m} and 70~\textmu{m}
(Fig.~\ref{fig2470obs}). The synthetic population shown corresponds to
$r\sbs{min}\approx 20$~AU, $r\sbs{max}\approx 120$~AU and $M\sbs{min} <
0.01$~$M_\oplus$, $M\sbs{max}\approx 30$~$M_\oplus$. Here, the radial range is
needed to cover the range of colors, i.e. the ratios between the excess
emissions at the two wavelengths. The mass range is needed to cover the observed
range of excess, especially for younger disks at 70~\textmu{m}.

Analyses of Spitzer detections might indicate a statistically significant
increase of both 24 and 70~\textmu{m} fluxes at ages between
a few tens of Myr to a few hundreds of Myr.
(e.g., J.~M. Carpenter et al., in prep.), which can only be marginally seen in
our sample (Fig.~\ref{fig2470obs}).
It is hypothesized that this feature
is caused either by an increased dust production due to delayed stirring by
growing planets or by events 
similar to the late heavy bombardment in the solar system.
Such effect could only be studied with an improved version of our analytic
model or with the numerical one.

\begin{figure}[t!]
  \includegraphics{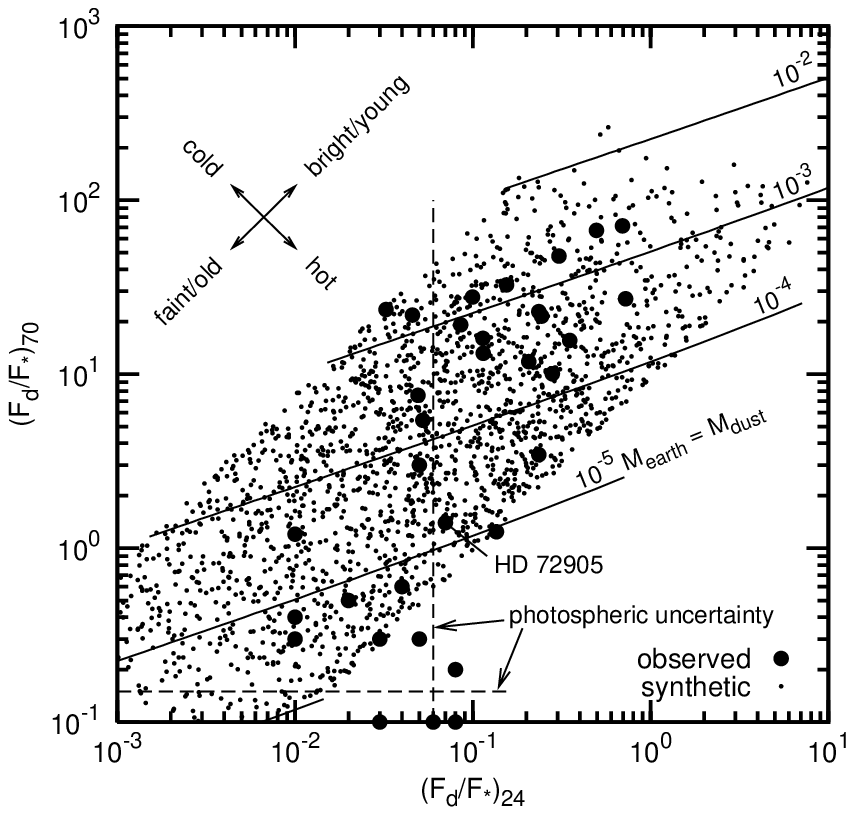}\\
  \includegraphics{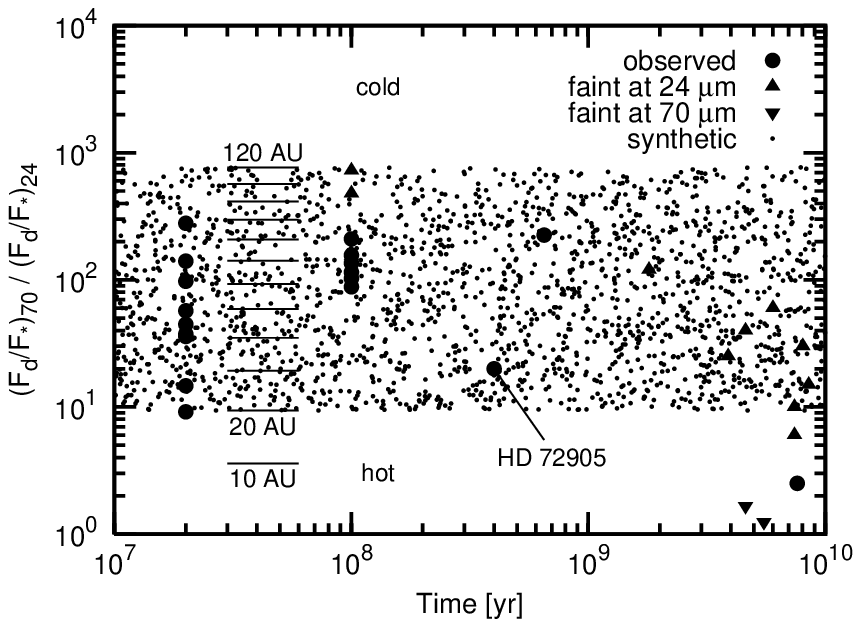}\\
  \caption{Relation between fluxes at 24 and 70~\textmu{m} versus time.
    The synthesized population (\textsl{small dots}) is compared to the
    observed one (\textsl{big dots and triangles}).
    The average photospheric uncertainty for both filters is marked by dashed
    lines in the upper panel. Excesses below those limits in either of the
    two filters are marked by triangles in the lower panel. In addition, the upper panel
    shows lines of equal dust mass and the lower panel gives the ring radii corresponding to the
    colors.
    \label{fig24/70}}
\end{figure}
The distribution of disk colors is more difficult to
reproduce. Fig.~\ref{fig24/70} shows a significant abundance of
fainter but warmer disks in an area that is not covered by the synthetic
population. One explanation would be that the upper mass limit is a function of
radial distance, and that the innermost disks tend to be less massive and less
luminous, from the very beginning. In addition, the lower panel of that
Fig.~\ref{fig24/70} shows a trend towards higher effective temperatures for
higher ages, which is difficult to understand. Indeed, as long as faint close-in
disks are observed around older stars, one would expect ever brighter disks, and therefore more
numerous detections of disks at the same distances
around younger stars.
Furthermore, the trend in question contradicts to the results by
\citet{najita-williams-2005}, who found no significant correlation 
between the disk radii and ages.
Most likely, the discrepancy is only caused by uncertainties of the 
measured excesses at 24~\textmu{m}. \citet{bryden-et-al-2006} report that the average
photometric accuracy in that filter band is only as good as $1 \sigma_{24}
= 6$\%\ due to stellar photosphere fitting errors and flat-field uncertainties.
Therefore, excesses below those 6\%\ of the photospheric emission cannot be
considered as significant. For 70~\textmu{m}, \citet{bryden-et-al-2006}
state $1 \sigma_{70} \approx 15$\%. Both limits are shown in the upper
panel of Fig.~\ref{fig24/70}.

In Figs.~\ref{fig2470obs} and \ref{fig24/70}, there is one particular system
directly labeled. That system, HD~72905, was observed to show significant excess
emission not only at 24 and 70~\textmu{m}, but also in the spectral
ranges
8--13 and 30--34~\textmu{m} of the Spitzer/IRS instrument
\citep{beichman-et-al-2006a}. The presence of two dusty regions was
suggested: one exozodiacal at 0.03--0.43~AU and one around 14~AU.
From the excess at 8--13~\textmu{m}, \citet{wyatt-et-al-2007a}
inferred the dust population in HD~72905 to be transient because the observed
fractional luminosity is above the maximum expected for a system of
300--400~Myr. As long as only 24 and 70~\textmu{m} are considered,
the HD~72905 dust does not seem particularly hot or bright, although
it is among the hotter disks.

At this point, it is interesting to compare our results to those of
\citet{wyatt-et-al-2007b}. Both analytic approaches aim at explaining and
reproducing the observations. Our model is different from theirs in that
we take into account the size dependence of the critical specific energy
as well as the transition from a ``primordial'' size distribution of planetesimals to the
one set up by a collisional cascade. 
The amount of dust in their model is determined, from the very beginning, by the
rather
long collisional timescales of objects of tens of kilometers, so that the
collisional evolution is much slower. This can be seen from the equations: ``1+'' in the 
denominator of Eq.~(\ref{eqMOverTime}) causes the mass to stay almost at the initial
level for a long time, before the system reaches the $t^{-1}$
decay. In our model, although the mass decay is asymptotically slower
($t^\xi$ with $\xi \approx -0.3 \ldots -0.4$),
it sets up very quickly, namely on collisional timescales of objects with minimum binding 
energy ($s_b \sim 100$ meters).
Therefore, we would expect the model by \citet{wyatt-et-al-2007b}
to show significantly larger excesses at ages considered, if all other parameters were
comparable. This, however, is not the case.
\citet{wyatt-et-al-2007b} assumed a much
weaker material in their collisional prescription.
Their $Q\sbs{D}^*=300$~J/kg at an object radius of 30~km ($D\sbs{c} = 60$~km)
is by more than two orders of magnitude below the
values we use in Eq.~(\ref{eqQD}). As $\tau \propto Q\sbs{D}^{q\sbs{p}-1}$ in Eq.~(\ref{eqtaured}), 
their collisional timescales are shorter and their evolution faster, too.
Besides the material strength, the difference in the assumed effective
eccentricities --- $e=0.05$ in their model against $e\sbs{max}/2 = 0.15$ in
ours~--- causes another factor of roughly 10 in the collisional timescales,
according to Sect.~\ref{secEcc}.
All the differences listed happen to nearly compensate each other.
As a net result, the excesses predicted by Wyatt's et al. and our models
are comparable with each other (see also Fig.~\ref{fig_fmax}), being in reasonable agreement with the observed ones.

\section{SUMMARY AND CONCLUSIONS}
\label{secConclusions}

We investigated the long-term evolution of debris disks around solar-type (G2V) stars.
Firstly, we performed numerical simulations with our collisional code.
Secondly, the numerical results
were supplemented by, and interpreted through, a new analytic model.
The latter is similar to, and builds up on, the model developed earlier by
\citet{wyatt-et-al-2007a}, but extends it in several important directions.
It naturally includes
the transition from the ``primordial'' size distribution of left-over 
planetesimals, set up at their agglomeration phase, to the size distribution
established by the collisional cascade. Further, it lifts the assumption
that the critical specific energy needed for disruption is constant
across the full range of sizes, from dust
to the largest planetesimals. With these improvements, a good agreement between the
numerics and analytics is achieved.

We draw the following conclusions:
\begin{enumerate}
  \item
    The timescale of the collisional evolution is inversely proportional to the initial disk mass.
    For example, halving the total mass doubles all collisional timescales.
    This rule is valid for systems where collisions are the only loss mechanism of particles
    and only as long as $\beta$-meteoroids are unimportant for the collisional
    budget.
  \item
    Numerics and analytics consistently yield a $\tau\propto  r^{4.3}$ 
    dependence of the timescale of the collisional evolution on the radial
    distance.
  \item
    Numerical simulations show that the collisional timescale
    varies with the average eccentricity of dust parent
    bodies as $\tau \propto e^{-2.3}$. The analytic
    approach suggests a somewhat weaker dependence, $\tau\propto e^{-5/3}$.
  \item
    An evolving three-slope size distribution is proposed to
    approximate the numerical results. The biggest objects are still distributed
    primordially, with a slope $q\sbs{p}$. The objects below a certain transitional size
    are already reprocessed by collisions and thus have a quasi-steady-state
    size distribution, determined by their self-gravity (for intermediate-sized
    objects, slope $q\sbs{g}$) or by material strength (for smallest objects, slope $q\sbs{s}$).
    That transitional size corresponds to the largest objects
    for which the collisional lifetime is still shorter than the age of the system.
    The transitional size increases with time, meaning that ever-larger planetesimals
    get involved into the collisional cascade.
  \item
    At actual ages of debris disks, $\sim$10~Myr to $\sim$10~Gyr,
    the decay of the dust mass and the total disk mass 
    follow {\em different} laws. The reason is that, in all conceivable debris disks,
    the largest planetesimals have longer collisional lifetimes than the system's age,
    and therefore did not have enough time to reach collisional equilibrium.
    If the system were let to evolve for sufficiently long time,
    both dust mass and disk mass would start to follow $t^{-1}$. However, this requires
    time spans of much longer than 10~Gyr.
  \item
    The loss rate of the dust mass, and the decay rate of fractional luminosity,
    primarily depend
    on the difference between the slope $q\sbs{p}$ of the ``primordial'' size distribution
    of largest planetesimals and the slope $q\sbs{g}$ of the size distribution of
    somewhat smaller, yet gravity-dominated, planetesimals that already underwent
    sufficient collisional evolution.
    With ``standard'' values of $q\sbs{p}$ and $q\sbs{g}$, the dust mass and the thermal fluxes
    follow approximately $t^\xi$ with $\xi = -0.3\ldots -0.4$.
  \item
    Specific decay laws of the total disk mass and the dust mass largely
    depend on a few model parameters.
    Most important are: the critical fragmentation energy $Q\sbs{D}^*$ as a
    function of size,
    the slope of the ``primordial'' size distribution of planetesimals $q\sbs{p}$
    and their maximum size $s\sbs{max}$,
    and the characteristic eccentricity $e$ and inclination $I$ of
    planetesimals.
  \item 
    The property that the maximum possible dust luminosity for a given age does
    not depend on the initial disk mass, established by \citet{wyatt-et-al-2007a},
    is only valid in cases of very rapid collisional evolution, i.e. in closer-in
    or dynamically very hot disks. For most of the systems 
    at ages $<10$~Gyr, an increase of the initial disk mass leads to an increase
    of the dust luminosity, unless that initial mass is assigned extreme
    values, incompatible with our understanding of planetesimal disks.
  \item
    Assuming standard material prescriptions and disk masses and extents,
    a synthetic population of disks generated with our analytic model generally
    agrees with the observed statistics of 24 and 70~\textmu{m} fluxes
    versus age.
    Similarly, the synthetic [24]-[70] colors are consistent with the 
    observed disk colors.
\end{enumerate}

As every model, our numerical model makes a number of general simplifying
assumptions; the analytic one imposes further simplifications:
\begin{itemize}
\item The collisional evolution is assumed to be smooth and
unperturbed. 
Singular episodes like the aftermath of giant break-ups or special periods of
the dynamical evolution such as the
late heavy bombardment are not included.
\item
Effects of possible perturbing planets are taken into account
only indirectly: through the eccentricities of planetesimals (dynamical excitation)
and confinement of planetesimal belts (truncation of disks).
Further effects such as resonant trapping or ejection of material by planets are neglected.
\item
We only consider disruptive collisions. This is a reasonable approximation
for disks that are sufficiently ``hot'' dynamically. However, cratering collisions
become important when the relative velocities are insufficient for
disruption to occur.
\item
Neither dilute disks under the regime of Poynting-Robertson drag nor
very dense disks with collisional timescales shorter than orbital timescales
and with avalanches \citep{grigorieva-et-al-2007} are covered by the present
work.
\item
Explaining the initial conditions or deriving them from the dynamical 
history of the systems at early stages of planetesimal and planetary accretion
was out of the scope of this paper. Correlations between disk masses,
disk radii, and the presence of planets, for example, were not considered,
although they might alter the scalings we found here.
\end{itemize}

Despite these limitations, our models reproduce, in essential part,
the observed evolution of dust in debris disks. We hope that they may serve
as a starting point for in-depth studies that will certainly be undertaken in the
future, motivated by questions that remain unanswered, as well as by new data
expected from ongoing and planned observational programs.

\acknowledgements

We wish to thank Jean-Fran{\c c}ois Lestrade, Philippe Th\'ebault, and Mark Wyatt for fruitful discussions
and Amaya Moro-Mart{\'i}n for useful review comments.
This research has been partly funded by the Deutsche
Forschungsgemeinschaft (DFG), project Kr 2164/5-1.



\end{document}